\documentstyle[epsf]{mn}

\title[Peculiar motions of early-type galaxies -- II]
{The peculiar motions of early-type galaxies in two distant regions -- II.
The spectroscopic data} 

\author[Wegner et al.]{
{\LARGE \rm Gary Wegner$^1$, Matthew Colless$^2$, R.\ P.\ Saglia$^3$,
Robert K. McMahan Jr$^{4,5}$,} \vspace*{6pt} \\ 
{\LARGE \rm Roger L.\ Davies$^6$, David Burstein$^7$ and Glenn Baggley$^6$}  
\vspace*{3pt} \\ 
$^1$Department of Physics and Astronomy, 6127 Wilder Laboratory,
Dartmouth College, Hanover, NH 03755-3528, U.S.A.  \\
$^2$Mount Stromlo and Siding Spring Observatories, The Australian 
National University, Weston Creek, ACT 2611, Australia \\
$^3$Universit\"{a}ts-Sternwarte M\"{u}nchen, Scheinerstra{\ss}e 1, 
D-81679 M\"{u}nchen, Germany \\
$^4$Dept of Physics and Astronomy, University of North Carolina,
CB 3255 Phillips Hall, Chapel Hill, NC 27599-3255, U.S.A. \\
$^5$P.O. Box 14026, McMahan Research Laboratories, 
79 Alexander Drive, Research Triangle, NC 27709, U.S.A. \\
$^6$Department of Physics, South Road, Durham DH1 3LE, United Kingdom \\
$^7$Department of Physics and Astronomy, Box 871054, Arizona State
University, Tempe, AZ 85287-1504, U.S.A. }

\date{Accepted ---. Received ---; in original form ---.}

\newcommand{\etal}{\hbox{et~al.}}
\newcommand{\ie}{\hbox{i.e.}}
\newcommand{\eg}{\hbox{e.g.}}

\newcommand{\Mpc}{\hbox{\,h$^{-1}$\,Mpc}}
\newcommand{\kpc}{\hbox{\,h$^{-1}$\,kpc}}
\newcommand{\kms}{\hbox{\,km\,s$^{-1}$}}

\newcommand{\mgb}{\hbox{Mg$b$}}
\newcommand{\mgbp}{\hbox{Mg$b^\prime$}}
\newcommand{\mgtwo}{\hbox{Mg$_2$}}
\newcommand{\mgsig}{\hbox{Mg--$\sigma$}}
\newcommand{\mgbpsig}{\hbox{\mgbp--$\sigma$}}
\newcommand{\mgtwosig}{\hbox{\mgtwo--$\sigma$}}
\newcommand{\n}{\phantom{0}}
\newcommand{\gs}
           {\mathrel{\hbox{\rlap{\hbox{\lower4pt\hbox{$\sim$}}}\hbox{$>$}}}}
\newcommand{\ls}
           {\mathrel{\hbox{\rlap{\hbox{\lower4pt\hbox{$\sim$}}}\hbox{$<$}}}}

\newcommand{\plotone}[1]
           {\centering \leavevmode \epsfxsize=\columnwidth \epsfbox{#1}}
\newcommand{\plotfull}[1]
           {\centering \leavevmode \epsfxsize=\textwidth \epsfbox{#1}}

\begin{document}

\maketitle

\begin{abstract}
We present the spectroscopic data for the galaxies studied in the EFAR
project, which is designed to measure the properties and peculiar
motions of early-type galaxies in two distant regions. We have obtained
1319 spectra of 714 early-type galaxies over 33 observing runs on 10
different telescopes. We describe the observations and data reductions
used to measure redshifts, velocity dispersions and the \mgb\ and
\mgtwo\ Lick linestrength indices. Detailed simulations and
intercomparison of the large number of repeat observations lead to
reliable error estimates for all quantities. The measurements from
different observing runs are calibrated to a common zeropoint or scale
before being combined, yielding a total of 706 redshifts, 676 velocity
dispersions, 676 \mgb\ linestrengths and 582 \mgtwo\ linestrengths. The
median estimated errors in the combined measurements are $\Delta
cz$=20\kms, $\Delta\sigma/\sigma$=9.1\%, $\Delta\mgb/\mgb$=7.2\% and
$\Delta\mgtwo$=0.015~mag. Comparison of our measurements with published
datasets shows no systematic errors in the redshifts or velocity
dispersions and only small zeropoint corrections to bring our
linestrengths onto the standard Lick system. We have assigned galaxies
to physical clusters by examining the line-of-sight velocity
distributions based on EFAR and ZCAT redshifts, together with the
projected distributions on the sky. We derive mean redshifts and
velocity dispersions for these clusters, which will be used in
estimating distances and peculiar velocities and to test for trends in
the galaxy population with cluster mass. The spectroscopic parameters
presented here for 706 galaxies combine high quality data, uniform
reduction and measurement procedures, and detailed error analysis. They
form the largest single set of velocity dispersions and linestrengths
for early-type galaxies published to date.
\end{abstract}


\section{INTRODUCTION}
\label{sec:intro}

We are measuring the peculiar motions of galaxy clusters in the
Hercules-Corona Borealis (HCB) and Perseus-Pisces-Cetus (PPC) regions at
distances between 6000 and 15000\kms\ using the global properties of
elliptical galaxies. This study (the EFAR project) has as primary goals:
(i)~characterising the intrinsic properties of elliptical galaxies in
clusters by compiling a large and homogeneous sample with high-quality
photometric and spectroscopic data; (ii)~testing possible systematic
errors, such as environmental dependence, in existing elliptical galaxy
distance estimators; (iii)~deriving improved distance estimators based
on a more comprehensive understanding of the properties of ellipticals
and how these are affected by the cluster environment; and
(iv)~determining the peculiar velocity field in regions that are
dynamically independent of the mass distribution within 5000\kms\ of our
Galaxy in order to test whether the large-amplitude coherent flows seen
locally are typical of bulk motions in the universe.

The background and motivation of this work are discussed in Paper~I of
this series (Wegner \etal\ 1996), which also describes in detail the
choice of regions to study, the sample of clusters and groups, and the
selection procedure and selection functions of the programme galaxies.
In earlier papers we reported the photoelectric photometry for 352
programme galaxies which underpins the transformation of our CCD data to
the standard $R$ magnitude system (Colless \etal\ 1993), and described
our technique for correcting for the effects of seeing on our estimates
of length scales and surface brightnesses (Saglia \etal\ 1993). This
paper (Paper~II) describes the spectroscopic observations and gives
redshifts, velocity dispersions and linestrength indices for the
programme galaxies. The CCD imaging observations of these galaxies, and
their photometric parameters, are described in Paper~III (Saglia \etal\
1997), while descriptions of the profile fitting techniques used to
determine these parameters (along with detailed simulations establishing
the uncertainties and characterising the systematic errors) are given in
Paper~IV (Saglia \etal\ 1997). The \mgsig\ relation and its implications
are discussed in Paper~V (Colless \etal\ 1998). Subsequent papers in the
series will explore other intrinsic properties of the galaxies and their
dependence on environment, derive an optimal distance estimator, and
discuss the peculiar motions of the clusters in each of our survey
regions and their significance for models of the large-scale structure
of the universe.

The structure of the present paper is as follows. In \S\ref{sec:obsvns}
we describe the observations and reductions used in obtaining the 1319
spectra in our dataset (1250 spectra for 666 programme galaxies and 69
spectra for 48 calibration galaxies) and discuss the quality of the
data. We explain the techniques by which redshifts, velocity dispersions
and linestrength indices were estimated from the spectra in
\S\ref{sec:analysis}, including the various corrections applied to the
raw values. In \S\ref{sec:results} we describe the method used to
combine data from different runs and evaluate the internal precision of
our results using the large number of repeat measurements in our
dataset. We then give the final values of the spectroscopic parameters
for each galaxy in our sample: we have redshifts for 706 galaxies,
dispersions and \mgb\ linestrengths for 676 galaxies and \mgtwo\
linestrengths for 582 galaxies. We compare our results to previous
studies in the literature to obtain external estimates of our random and
systematic errors. In \S\ref{sec:clusass} we combine our redshifts with
those from ZCAT in order to assign sample galaxies to physical clusters,
and to estimate the mean redshifts and velocity dispersions of these
clusters. Our conclusions are summarised in \S\ref{sec:conclude}.

This paper presents the largest and most homogeneous sample of velocity
dispersions and linestrengths for elliptical galaxies ever obtained. The
precision of our measurements is sufficiently good to achieve the goal
of measuring distances via the Fundamental Plane out to 15000\kms.

\section{OBSERVATIONS}
\label{sec:obsvns}

The spectroscopic observations for the EFAR project were obtained over a
period of seven years from 1986 to 1993 in a total of 33 observing runs
on 10 different telescopes. In this section we describe the
spectroscopic setups, the observing procedures, the quality of the
spectra and the data-reduction techniques. Further detail on these
points is given by Baggley (1996).

\subsection{Spectroscopic Setups}
\label{ssec:setups}

Table~\ref{tab:obsruns} gives the spectroscopic setup for each run,
including the run number, date, telescope, spectrograph and detector,
wavelength range, spectral dispersion (in \AA/pixel), effective
resolution (in \kms), and the effective aperture size. Note that two
runs (116 and 130) produced no useful data and are included in
Table~\ref{tab:obsruns} only for completeness. Three runs utilised fibre
spectrographs: runs 127 and 133 used Argus on the CTIO 4m and run 131
used MEFOS on the ESO 3.6m. All the other runs employed longslit
spectrographs, mostly on 2m-class telescopes (MDM Hiltner 2.4m, Isaac
Newton 2.5m, Kitt Peak 2.1m, Siding Spring 2.3m, Calar Alto 2.2m)
although some 4m-class telescopes were also used (Kitt Peak 4m, William
Herschel 4m, the MMT).

\begin{table*}
\centering
\caption{The Spectroscopic Observing Runs}
\label{tab:obsruns}
\begin{tabular}{lccllccclr}
Run & Date & Tele- $^a$ & Spectrograph & Detector &
$\lambda\lambda$ & $\Delta\lambda$ $^b$ & 
$\sigma_i$ $^c$ & Aperture $^d$ & $N^e$ \\
~\# & & scope & + grating & & (\AA) & (\AA/pix) & (km/s) & (arcsec) & 
\vspace*{6pt} \\ 
101     & 86/12 & MG24 & MarkIIIa+600B   & GEC       & 4912--6219 & 2.27 & 145 & 1.9$\times$10\% &  22 \\
102     & 87/03 & MG24 & MarkIIIa+600B   & RCA       & 4787--6360 & 3.10 & 145 & 1.9$\times$10\% &  58 \\
103     & 87/05 & MG24 & MarkIIIa+600B   & RCA       & 4809--6364 & 3.07 & 145 & 1.9$\times$10\% &  37 \\
104     & 88/04 & MG24 & MarkIIIa+600V   & RCA       & 5025--6500 & 2.90 & 125 & 1.9$\times$10\% &  12 \\
105     & 88/06 & MG24 & MarkIIIa+600V   & RCA       & 5055--6529 & 2.90 & 125 & 1.9$\times$10\% &  37 \\
106     & 88/09 & MG24 & MarkIIIa+600V   & Thompson  & 5041--6303 & 2.21 & 130 & 1.9$\times$10\% &  23 \\
107     & 88/10 & MG24 & MarkIIIa+600V   & RCA       & 5048--6522 & 2.90 & 130 & 1.9$\times$10\% &  27 \\
108     & 88/07 & MMTB & BigBlue+300B    & Reticon   & 3700--7200 & 1.14 & 135 & 2.5             &  10 \\
109     & 88/11 & KP4M & RC+UV-Fast+17B  & TI2       & 4890--5738 & 1.07 & 100 & 2.0$\times$3.9  & 104 \\
110     & 88/11 & KP2M & GoldCam+\#240   & TI5       & 4760--5879 & 1.52 & 105 & 2.0$\times$3.9  &  72 \\
111     & 88/11 & MMTB & BigBlue+300B    & Reticon   & 3890--7500 & 1.34 & 135 & 2.5             &  20 \\
112     & 89/04 & MG24 & MarkIIIb+600V   & RCA       & 5066--6534 & 2.91 & 130 & 1.7$\times$10\% &  34 \\
113     & 89/06 & MG24 & MarkIIIb+600V   & TI4849    & 5126--6393 & 2.18 & 150 & 2.4$\times$10\% &  23 \\
114     & 89/06 & MMTB & BigBlue+300B    & Reticon   & 3890--7500 & 1.34 & 135 & 2.5             &  12 \\
115     & 89/08 & WHT4 & ISIS-Blue+R600B & CCD-IPCS  & 4330--4970 & 0.45 &\n95 & 2.0$\times$3.9  &   8 \\
116$^f$ & 89/10 & MMTR & RedChannel+600B & TI        & 4750--5950 & 1.50 & 125 & 1.5$\times$10\% &   7 \\
117     & 89/10 & MG24 & MarkIIIb+600V   & Thompson  & 5031--6300 & 2.21 & 130 & 1.7$\times$10\% &  61 \\
118     & 89/11 & MG24 & MarkIIIb+600V   & RCA       & 5018--6499 & 2.89 & 170 & 1.7$\times$10\% &  14 \\
119     & 90/05 & MMTR & RedChannel+600B & TI        & 4750--5950 & 1.50 & 125 & 1.5$\times$10\% &  17 \\
120     & 90/10 & IN25 & IDS+235mm+R632V & GEC6      & 4806--5606 & 1.46 & 100 & 1.9$\times$7.2  &  40 \\
121     & 91/05 & IN25 & IDS+235mm+R632V & GEC3      & 4806--5603 & 1.46 & 100 & 1.9$\times$7.2  &  87 \\
122     & 91/10 & MG24 & MarkIIIb+600V   & Thompson  & 5018--6278 & 2.20 & 125 & 1.7$\times$10\% &  43 \\
123     & 91/11 & IN25 & IDS+235mm+R632V & GEC3      & 4806--5603 & 1.46 & 100 & 1.9$\times$7.2  &  29 \\ 
124     & 92/01 & MG24 & MarkIIIb+600V   & Thompson  & 5038--6267 & 2.15 & 125 & 1.7$\times$10\% &  35 \\
125     & 92/06 & MG24 & MarkIIIb+600V   & Loral     & 4358--7033 & 1.31 & 125 & 1.7$\times$10\% &  57 \\
126     & 92/06 & CA22 & B\&C~spec+\#7   & TEK6      & 4800--6150 & 1.40 & 100 & 5.0$\times$4.2  &  39 \\
127$^g$ & 92/09 & CT4M & Argus+KPGL\#3   & Reticon II& 3877--6493 & 2.19 & 145 & 1.9             & 199 \\
128     & 93/05 & MG24 & MarkIIIb+600V   & Loral     & 5090--7050 & 1.40 & 105 & 1.7$\times$10\% &  24 \\
129     & 93/06 & MG24 & MarkIIIb+600V   & TEK       & 4358--6717 & 2.31 & 135 & 1.7$\times$10\% &   3 \\
130$^f$ & 93/06 & SS23 & DBS-blue+600B   & PCA       & 5015--5555 & 0.80 &\n85 & 2.0$\times$10\% &   0 \\ 
131$^g$ & 93/10 & ES36 & MEFOS+B\&C+\#26 & TEK512CB  & 4850--5468 & 1.22 & 105 & 2.6             & 128 \\
132     & 93/10 & SS23 & DBS-blue+600B   & Loral     & 4820--5910 & 1.10 &\n80 & 2.0$\times$10\% &  14 \\ 
133$^g$ & 93/10 & CT4M & Argus+KPGL\#3   & Reticon II& 3879--6485 & 2.19 & 145 & 1.9             & 193 \\
\end{tabular}\vspace*{6pt}
\parbox{\textwidth}{
$^a$ Telescopes: MG24=MDM 2.4m; KP4M/KP2M=KPNO 4m/2m;
WHT4=William Herschel 4m (LPO); IN25=Isaac Newton 2.5m (LPO);
MMTB/MMTR=MMT (blue/red); CA22=Calar Alto 2.2m; CT4M=Cerro Tololo
4m; SS23=Siding Spring 2.3m; ES36=ESO 3.6m. \\
$^b$ Spectral dispersion in \AA/pixel. \\
$^c$ Instrumental resolution ($\sigma$, not FWHM) in \kms, as
determined from the cross-correlation analysis calibration curves (see
\S\ref{ssec:czsig}). \\
$^d$ The aperture over which the galaxy spectrum was extracted:
diameter for circular apertures and fibres, width$\times$length for
rectangular slits (10\% means the spectrum was extracted out to the
point where the luminosity had fallen to 10\% of its peak value). \\
$^e$ The number of spectra taken in the run. \\
$^f$ These runs produced no useful data. \\
$^g$ These runs used fibre spectrographs. }
\end{table*}

The spectra from almost all runs span at least the wavelength range
5126--5603\AA, encompassing the MgI\,$b$ 5174\AA\ band and the FeI
5207\AA\ and 5269\AA\ features in the restframe for galaxies over the
sample redshift range $cz$$\approx$6000--15000\kms. The exceptions are
the spectra from runs 115 and 131. Run 115 comprises 8 spectra obtained
at the WHT with the blue channel of the ISIS spectrograph which have a
red wavelength limit of 4970\AA\ (\ie\ including H$\beta$ but not
\mgb). Since we have spectra for all these galaxies from other runs we
do {\em not} use the redshifts and dispersions from run 115. Run 131
comprises 128 spectra obtained at the ESO 3.6m with the MEFOS fibre
spectrograph to a red limit of 5468\AA, including \mgb\ and FeI 5207\AA\
over the redshift range of interest, but not FeI 5269\AA\ beyond
$cz$$\approx$11000\kms. For most of the runs the spectra also encompass
H$\beta$, and several span the whole range from CaI~H+K 3933+3969\AA\ to
NaI~D 5892\AA.

The effective instrumental resolution of the spectra, $\sigma_i$, was
measured from the autocorrelation of stellar template spectra (see
\S\ref{ssec:czsig} below), and ranged from 80 to 170\kms, with a median
value of 125\kms. Both longslit and circular entrance apertures were
used. Slits were typically 1.7--2.0~arcsec wide and the spectra were
extracted to the point where the galaxy fell to about 10\% of its peak
value. Circular apertures (in the fibre spectrographs and the MMT Big
Blue spectrograph) were between 1.9 and 2.6~arcsec in diameter.  Further
details of the observing setup for each telescope/instrument combination
are given in Appendix~A.

\subsection{Observing Procedures}
\label{ssec:obsproc}

The total integration times on programme galaxies varied considerably
depending on telescope aperture, observing conditions and the magnitude
and surface brightness of the target (our programme galaxies have R band
total magnitudes in the range 10--16). On 2m-class telescopes (with
which the bulk of the spectroscopy was done), exposure times were
usually in the range 30--60~min, with a median of 40~min; on 4m-class
telescopes, exposure times were generally 15--20~min (up to 60~min for
the faintest galaxies) with single-object slit spectrographs, but 60 or
120~min with the fibre spectrographs (where the aim was high $S/N$ and
completeness). Slit orientations were not generally aligned with galaxy
axes. The nominal goal in all cases was to obtain around 500
photons/\AA\ at \mgb, corresponding to a $S/N$ per 100\kms\ resolution
element of about 30. In fact our spectra have a median of 370
photons/\AA\ at \mgb, corresponding to a $S/N$ per 100\kms\ of 26 (see
\S\ref{ssec:quality}).

In each run several G8 to K5 giant stars with known heliocentric
velocities were observed. These `velocity standard stars' are used as
spectral templates for determining redshifts and velocity dispersions.
In observing these standards care was taken to ensure that the
illumination across the slit was uniform, in order both to remove
redshift zeropoint errors and to mimic the illumination produced by a
galaxy, thereby minimising systematic errors in velocity dispersion
estimates. This was achieved in various ways: by defocussing the
telescope slightly, by moving the star back and forth across the slit
several times, or by trailing it up and down the slit. Such procedures
were not necessary for standards obtained with fibre spectrographs, as
internal reflections in the fibres ensure even illumination of the
spectrograph for all sources. Very high $S/N$ (typically $>$10,000
photons/\AA) were obtained in order that the stellar templates did not
contribute to the noise in the subsequent analysis.

The normal calibration exposures were also obtained: bias frames,
flatfields (using continuum lamps internal to the spectrographs or
illuminating the dome) and spectra of wavelength calibration lamps
before and/or after each galaxy or star exposure. In general we did not
make use of spectrophotometric standards as fluxed spectra were not
necessary and we wished to minimise overheads as much as possible.

The calibration procedures were slightly different for the three large
datasets taken using fibre-fed spectrographs at CTIO (runs 127 and 133)
and ESO (run 131). Because of the need to calibrate the relative
throughput of the fibres in order to perform sky subtraction, fibre
observations always included several twilight sky flatfield
exposures. Each velocity standard star was observed through several
fibres by moving the fibres sequentially to accept the starlight.

\subsection{Reductions}
\label{ssec:reduce}

The reductions of both the longslit and fibre observations followed
standard procedures as implemented in the IRAF\footnote{IRAF is
distributed by the National Optical Astronomy Observatories which is
operated by the Association of Universities for Research in Astronomy,
Inc. under contract with the National Science Foundation.}, MIDAS and
Starlink Figaro software packages. We briefly summarise the main steps
in the reduction of our longslit and fibre data below; further details
can be found in Baggley (1996).

The first stage of the reductions, common to all observations, was to
remove the CCD bias using a series of bias frames taken at the start or
end of the night. These frames were median-filtered and the result,
scaled to the mean level of the CCD overscan strip, was subtracted from
each frame in order to remove both the spatial structure in the bias
pedestal and temporal variations in its overall level. We also took long
dark exposures to check for dark current, but in no case did it prove
significant. Subsequent reductions differed somewhat for longslit and
fibre observations.

For longslit data, the next step was the removal of pixel-to-pixel
sensitivity variations in the CCD by dividing by a sensitivity map. This
map was produced by median-filtering the flatfield exposures (of an
internal calibration lamp or dome lamp) and dividing this by a smoothed
version of itself (achieved by direct smoothing or 2D surface fitting)
in order to remove illumination variations in the `flat' field. If
necessary (because of a long exposure time or a susceptible CCD), cosmic
ray events were identified and interpolated over in the two-dimensional
image using either algorithmic or manual methods (or both).

The transformation between wavelength and pixel position in longslit
data was mapped using the emission lines in the comparison lamp
spectra. The typical precision achieved in wavelength calibration, as
indicated by the residuals of the fit to the calibration line positions,
was $\ls$\,0.1\,pixel, corresponding to 0.1--0.3\AA\ or 5--15\kms,
depending on the spectrograph setup (see Table~\ref{tab:obsruns}). The
spectra were then rebinned into equal intervals of $\log\lambda$ so that
each pixel corresponded to a fixed velocity interval, $\Delta v \equiv
c\Delta z = c(10^{\Delta\log\lambda}-1)$, chosen to preserve the full
velocity resolution of the data.

The final steps in obtaining longslit spectra are sky-subtraction and
extraction. The sky level was measured from two or more regions along
the slit sufficiently far from the target object to be uncontaminated by
its light. To account for variations in transmission along the slit, the
sky under the object was interpolated using a low-order curve fitted to
the slit illumination profile. A galaxy spectrum was then extracted by
summing along the profile, usually over the range where the object's
luminosity was greater than $\sim$10\% of its peak value, but sometimes
over a fixed width in arcsec (see Table~\ref{tab:obsruns}). Standard
star spectra were simply summed over the range along the slit that they
had been trailed or defocussed to cover.

For the fibre runs the individual object and sky spectra were extracted
first, using a centroid-following algorithm to map the position of the
spectrum along the CCD. The extraction algorithm fitted the spatial
profile of the fibre, in order to remove cosmic ray events and pixel
defects, and then performed a weighted sum over this fit out to the
points where the flux fell to $\sim$5\% of the peak value. Next, the
dome-illumination flatfield spectra were median-combined and a
sensitivity map for each fibre constructed by dividing each fibre's
flatfield spectrum by the average over all fibres and normalising the
mean of the result to unity. The pixel-to-pixel variations in the CCD
response were then removed by dividing all other spectra from that fibre
by this sensitivity map. Wavelength calibration was accomplished using
the extracted comparison lamp spectra, giving similar precision to the
longslit calibrations, and the object spectra were rebinned to a
$\log\lambda$ scale. Using the total counts through each fibre from the
twilight sky flatfield to give the relative throughputs, the several sky
spectra obtained in each fibre exposure were median-combined (after
manually removing `sky' fibres which were inadvertently placed on faint
objects). The resulting high-$S/N$ sky spectrum, suitably normalised to
each fibre's throughput, was then subtracted from each galaxy or
standard star spectrum.

The final step in the reductions for both longslit and fibre data was to
manually clean all the one-dimensional spectra of remaining cosmic ray
events or residual sky lines (usually only the 5577\AA\ line) by
linearly interpolating over affected wavelengths. 

\subsection{Spectrum Quality}
\label{ssec:quality}

\begin{figure*}
\plotfull{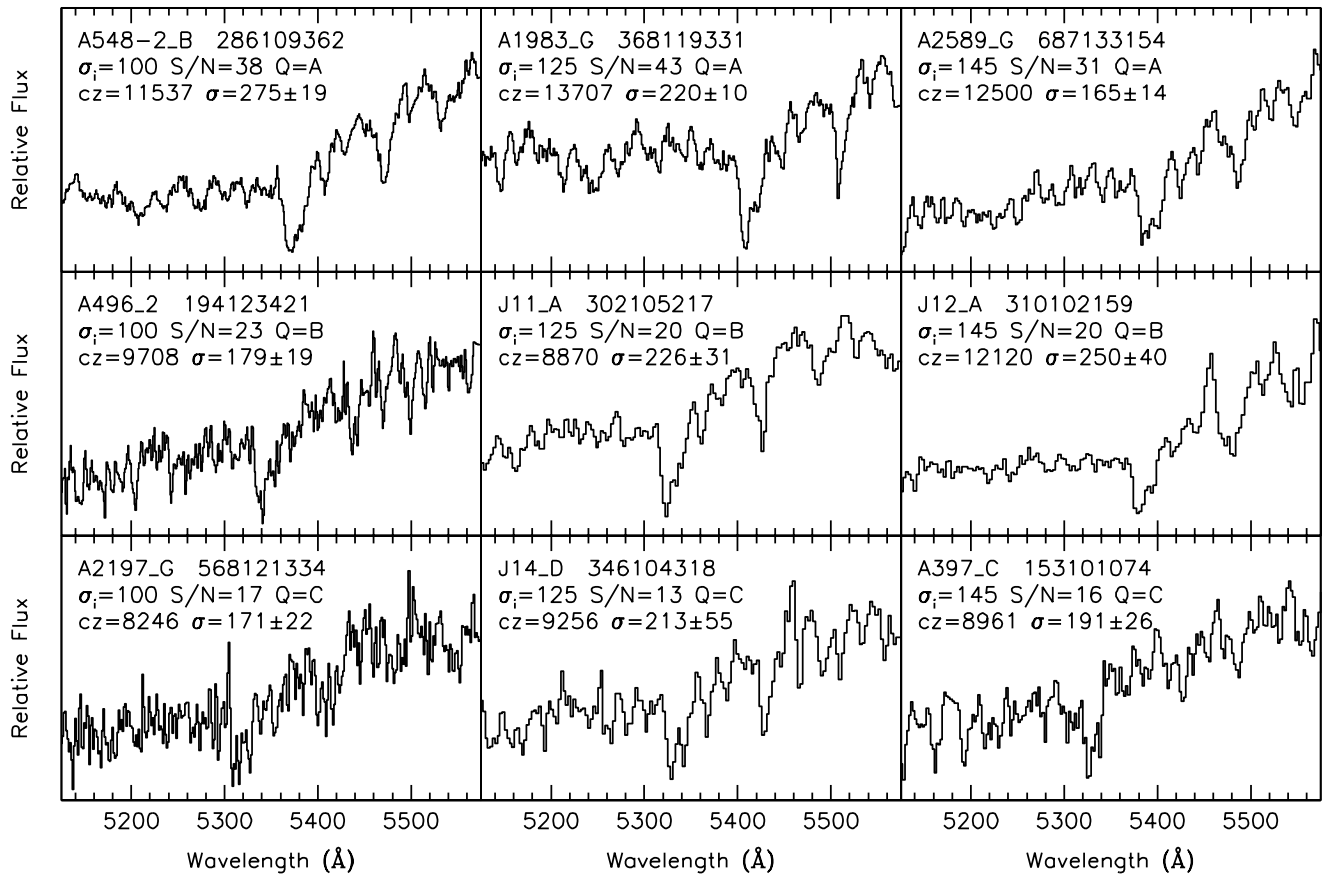}
\caption{Example spectra covering a range of quality classes and
instrumental resolutions: the top, middle and bottom rows are spectra
with quality classes A, B and C respectively; the left, central and
right columns are spectra with resolutions 100, 125 and 145\kms\
respectively. The label for each spectrum gives the galaxy name, the
GINRUNSEQ spectrum identifier, the instrumental resolution, the S/N and
quality class of the spectrum, the redshift, the dispersion and its
estimated error. Note that the panels show relative flux and have a
false zero for viewing convenience.
\label{fig:egspec}}
\end{figure*}

We have two methods for characterising the quality of our spectra. One
is a classification of the spectra into 5 quality classes, based on our
experience in reducing and analysing such data. Classes A and B indicate
that both the redshift and the velocity dispersion are reliable (with
class A giving smaller errors than class B); class C spectra have
reliable redshifts and marginally reliable dispersions; class D spectra
have marginally reliable redshifts but unreliable dispersions; class E
spectra have neither redshifts nor dispersions. The second method is
based on the $S/N$ ratio per 100\kms\ bin, estimated approximately from
the mean flux over the restframe wavelength range used to determine the
redshifts and dispersions (see \S\ref{ssec:czsig}) under the assumption
that the spectrum is shot-noise dominated. These two measures of
spectral quality are complementary: the $S/N$ estimate is objective but
cannot take into account qualitative problems which are readily
incorporated in the subjective classifications. Figure~\ref{fig:egspec}
shows example spectra covering a range of quality classes and
instrumental resolutions.

\begin{figure}
\plotone{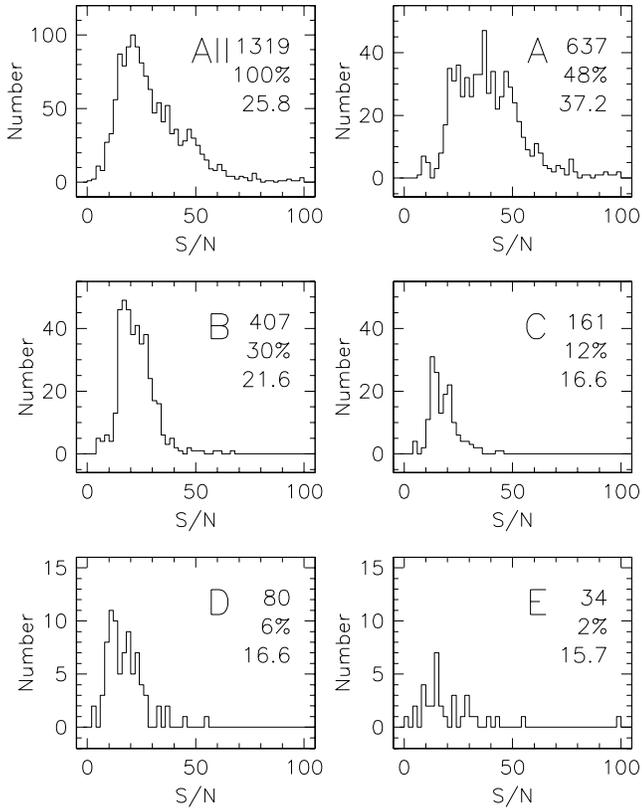}
\caption{The distribution of $S/N$ with quality class. For each class
the panels give the total number of spectra, the percentage of the
whole sample and the median $S/N$.
\label{fig:snrq}}
\end{figure}

Figure~\ref{fig:snrq} shows the $S/N$ distribution for the whole sample
and for each quality class individually, and gives the total number of
objects, the fraction of the sample and the median $S/N$ in each class.
For the whole sample, 39\% of the spectra have $S/N$$>$30, 70\% have
$S/N$$>$20, and 96\% have $S/N$$>$10. The two quality measures are
clearly correlated, in the sense that better-classed spectra tend to
have higher $S/N$. However there is also considerable overlap in the
$S/N$ range spanned by the different classes. This overlap has various
sources: (i)~factors other than $S/N$ which affect the quality of the
redshift and dispersion estimates, notably the available restframe
spectral range (which depends on both the spectrograph setup and the
redshift of the target) and whether the object has emission lines;
(ii)~errors in estimating the $S/N$ (\eg\ due to sky subtraction errors,
the neglect of the sky contribution in computing the $S/N$ for fainter
galaxies, or uncertainties in the CCD gain (affecting the conversion
from counts to photons); (iii)~subjective uncertainties in the quality
classification, particularly in determining the reliability of
dispersion estimates (\ie\ between classes B and C). Both ways of
determining spectral quality are therefore needed in order to estimate
the reliability and precision of the spectroscopic parameters we
measure.

\section{ANALYSIS}
\label{sec:analysis}

\subsection{Redshifts and Dispersions}
\label{ssec:czsig}

We derived redshifts and velocity dispersions from our spectra using the
{\tt fxcor} utility in IRAF, which is based on the cross-correlation
method of Tonry \& Davis (1979). We preferred this straightforward and
robust method to more elaborate techniques since it is well-suited to
the relatively modest $S/N$ of our spectra. We used a two-step
procedure, obtaining an initial estimate of the redshift using the whole
available spectrum and then using a fixed restframe wavelength range for
the final estimates of redshift and velocity dispersion. The procedure
was applied in a completely uniform manner to all the spectra in our
sample as far as differences in wavelength range and resolution would
allow.

The first step in the cross-correlation analysis is to fit and subtract
the continuum of each spectrum in order to avoid the numerical
difficulties associated with a dominant low-frequency spike in the
Fourier transform. In the first pass through {\tt fxcor} the continuum
shape was fitted with a cubic spline with the number of segments along
the spectrum chosen so that each segment corresponded to about
8000\kms. Each iteration of the fit excluded points more than
1.5$\sigma$ below or 3$\sigma$ above the previous fit. In this way we
achieved a good continuum fit without following broad spectral
features. We then apodised 10\% of the spectrum at each end with a
cosine bell before padding the spectrum to 2048 pixels with zeros.

This continuum-subtracted, apodised spectrum was then Fourier
transformed and a standard `ramp' filter applied. This filter is
described by 4 wavenumbers $(k_1,k_2,k_3,k_4)$, rising linearly from 0
to 1 between $k_1$ and $k_2$ and then falling linearly from 1 to 0
between $k_3$ and $k_4$. In the first pass these wavenumbers were chosen
to be $k_1$=4--8 and $k_2$=9--12 (tailored to remove residual power from
the continuum without affecting broad spectral features), and
$k_3$=$N_{pix}$/3 and $k_4$=$N_{pix}$/2 ($N_{pix}$ is the number of
pixels in original spectrum before it is padded to 2048 pixels; these
choices attenuate high-frequency noise and eliminate power beyond the
Nyquist limit at $N_{pix}$/2). The same procedures were also applied to
the spectrum of the stellar velocity standard to be used as a
template. The cross-correlation of the galaxy and stellar template was
then computed, and the top 90\% of the highest cross-correlation peak
fitted with a Gaussian in order to obtain a redshift estimate.

This procedure was repeated for every template from that run, and the
redshifts corrected to the heliocentric frame. Offsets in the velocity
zeropoint between templates, measured as the mean difference in the
redshifts measured with different templates for all the galaxies in the
run, were typically found to be $\ls$\,30\kms. These were brought into
relative agreement within each run by choosing the best-observed K0
template as defining the fiducial velocity zeropoint. Applying these
offsets brought the galaxy redshifts estimated from different templates
into agreement to within $\ls$\,3\kms. (The removal of run-to-run
velocity offsets is described below.) The mean over all templates then
gave the initial redshift estimate for the galaxy.

This initial redshift was then used to determine the wavelength range
corresponding to the restframe range $\lambda_{min}$=4770\AA\ to
$\lambda_{max}$=5770\AA. This range was chosen for use in the second
pass through {\tt fxcor} because: (i)~it contains the MgI\,$b$
5174\AA\ band, H$\beta$ 4861\AA\ and the FeI 5207\AA\ and 5269\AA\
lines, but excludes the NaI~D line at 5892\AA, which gives larger
velocity dispersions than the lines in the region of \mgb\ (Faber \&
Jackson 1976); (ii)~for redshifts up to our sample limit of
$cz$=15000\kms\ this restframe wavelength range is included in the
great majority of our spectra. The input for the second pass was thus
the available spectrum within the range corresponding to restframe
4770--5770\AA. All but two of our runs cover the restframe out to at
least 5330\AA\ for $cz$=15000\kms; the exceptions are run 115 (which
is not used for measuring dispersions) and run 131 (which reaches
restframe 5207\AA).

In the second pass through {\tt fxcor} we employed only minimal
continuum subtraction based on a 1- or 2-segment cubic spline fit,
preferring the better control over continuum suppression afforded by
more stringent filtering at low wavenumbers. After considerable
experimentation and simulation, we found that the best filter for
recovering velocity dispersions was a ramp with the same $k_3$ and
$k_4$ values as in the first pass, but with
$k_2$=$0.01(N_{max}-N_{min})$, where $N_{min}$ and $N_{max}$ are the
pixels corresponding to $\lambda_{min}$ and $\lambda_{max}$, and
$k_1$=0.75$k_2$. Again, the top 90\% of the highest cross-correlation
peak was fitted with a Gaussian. The position of this peak, corrected
for the motion of the template star and the heliocentric motion of the
earth relative to both the template and the galaxy, gave the final
redshift estimate.

\begin{figure}
\plotone{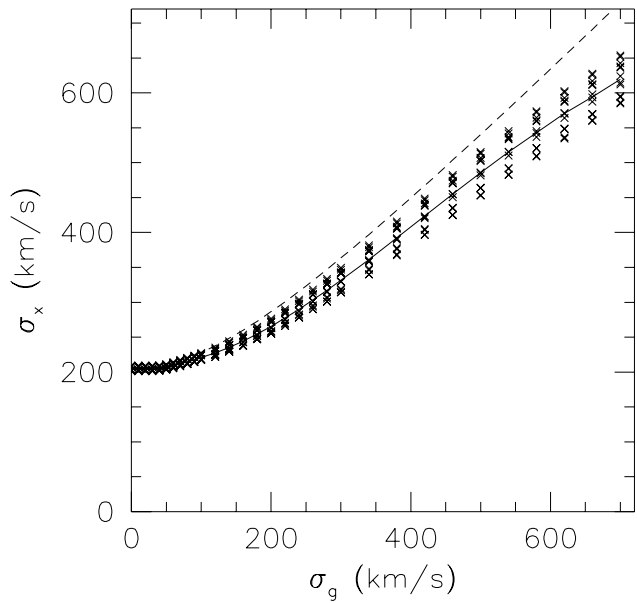}
\caption{A typical calibration curve showing the relation between the
width of the cross-correlation peak, $\sigma_x$, and the true velocity
dispersion of the galaxy, $\sigma_g$. The crosses are the individual
calibrations obtained by broadening each of the other templates in the
run and cross-correlating with the template being calibrated. The
solid curve is the calibration curve used, a series of linear segments
joining the median value of $\sigma_x$ at each calibrated value of
$\sigma_g$. The dashed curve is the theoretical relation when no
filtering is applied, $\sigma_x^2 = \sigma_g^2 + 2\sigma_i^2$, where
$\sigma_i$ is the instrumental resolution, in this case 145\kms. Note
that the calibration curve flattens for $\sigma_g<\sigma_i$,
indicating that the true dispersion becomes increasingly difficult to
recover as it drops below the instrumental resolution.
\label{fig:calib}}
\end{figure}

The galaxy's velocity dispersion, $\sigma_g$, is in principle related to
the dispersion of the Gaussian fitted to the cross-correlation peak,
$\sigma_x$, by $\sigma_x^2 = \sigma_g^2 + 2\sigma_i^2$ (where $\sigma_i$
is the instrumental resolution; Tonry \& Davis 1979). In practice this
relationship needs to be calibrated empirically because of the imperfect
match between the spectra of a broadened stellar template and a galaxy
and the effects of the filter applied to both spectra. The calibration
relation between $\sigma_x$ and $\sigma_g$ for a typical case is shown
in Figure~\ref{fig:calib} (see caption for more details). We estimate
the instrumental resolution for a given run from the mean value of the
calibration curve intercepts for all the templates in the run
($\sigma_i\approx\sigma_x/\sqrt{2}$ when $\sigma_g$=0); these are the
values listed in Table~\ref{tab:obsruns}.

The values of heliocentric radial velocity and velocity dispersion were
determined in this second pass through {\tt fxcor} for each galaxy
spectrum using all the templates in the same run. The final step is then
to combine the redshift and dispersion estimates from each template, as
summarised below.

For the redshifts the steps involved were as follows: (i)~Cases where
the ratio of cross-correlation function peak height to noise (the $R$
parameter defined by Tonry \& Davis 1979) was less than 2 were rejected,
as were cases that differed from the median by more than a few hundred
\kms. (ii)~The mean offset between the redshifts from a fiducial K0
template and each other template was used to shift all the redshifts
from the other template to the velocity zeropoint of the fiducial. These
offsets were typically $\ls$\,50\kms. (iii)~A mean redshift for each
galaxy was then computed from all the unrejected cases using 2-pass
2$\sigma$ clipping. (iv)~Any template which gave consistently discrepant
results was rejected and the entire procedure repeated. The scatter in
the redshift estimates from different templates after this procedure was
typically a few \kms.

A very similar procedure was followed in combining velocity dispersions
except that a scale factor rather than an offset was applied between
templates: (i)~Cases with $R$$<$4 were rejected.  (ii)~The mean ratio
between the dispersions from a fiducial K0 template and each other
template was used to scale all the dispersions from the other template
to the dispersion scale of the fiducial. These dispersion scales
differed by less than 5\% for 90\% of the templates.  (iii)~A mean
dispersion for each galaxy was then computed from all the unrejected
cases using 2-pass 2$\sigma$ clipping. (iv)~Any template with a scale
differing by more than 10\% from the mean was rejected as being a poor
match to the programme galaxies and the entire procedure was then
repeated. (Note that no significant correlation was found between scale
factor and spectral type over the range G8 to K5 spanned by our
templates.) The scatter in the dispersion estimates from different
templates after this procedure was typically 3--4\%.

Two corrections need to be applied to the velocity dispersions before
they are fully calibrated: (i)~an aperture correction to account for
different effective apertures sampling different parts of the galaxy
velocity dispersion profile, and (ii)~a run correction to remove
systematic scale errors between different observing setups. The latter
type of correction is also applied to the redshifts to give them a
common zeropoint. These two corrections are discussed below at
\S\ref{ssec:apcorr} and \S\ref{ssec:combruns} respectively.

\subsection{Linestrength Indices}
\label{ssec:indices}

Once redshifts and velocity dispersions were determined, linestrength
indices could also be measured using the prescription given by
Gonz\'{a}lez (1993). This is a refinement of the original `Lick' system
in which a standard set of bands was defined for measuring linestrength
indices for 11 features in the spectra of spheroidal systems (Burstein
\etal\ 1984). Gonz\'{a}lez (1993), Worthey (1993) and Worthey \etal\
(1994) describe how this system has been updated and expanded to a set of
21 indices. Here we measure both the \mgb\ and \mgtwo\ indices.

The feature bandpass for \mgb\ index is 5160.1--5192.6\AA, encompassing
the Mg~I triplet with components at 5166.6\AA, 5172.0\AA\ and
5183.2\AA. The continuum on either side of the absorption feature is
defined in bands covering 5142.6--5161.4\AA\ and 5191.4--5206.4\AA.
\mgb\ is an {\em atomic} index, and so is defined as the equivalent
width of the feature in {\AA}ngstroms,
\begin{equation}
\mgb = \int\,\left(1-\frac{S(\lambda)}{C(\lambda)}\right)\,d\lambda ~,
\label{eqn:mgbdef}
\end{equation}
where the integral is over the feature bandpass, $S(\lambda)$ is the
object spectrum and $C(\lambda)$ is the linear pseudo-continuum defined
by interpolating between two continuum estimates, taken at the midpoints
of the blue and red continuum bands to be the mean values of the
observed spectrum in those bands.

Closely related to \mgb\ is the \mgtwo\ index, for which the feature
bandpass is 5154.1--5196.6\AA\ and the continuum bands are
4895.1--4957.6\AA\ and 5301.1--5366.1\AA. This index measures both the
Mg~I atomic absorption and the broader MgH molecular absorption
feature. \mgtwo\ is a {\em molecular} index, and so is defined as the
mean ratio of flux to local continuum in magnitudes,
\begin{equation}
\mgtwo = -2.5\log_{10}\left(\frac{\int\,S(\lambda)/C(\lambda)\,d\lambda}
                                 {\Delta\lambda}\right) ~,
\label{eqn:mg2def}
\end{equation}
where the integral is over the \mgtwo\ feature bandpass,
$\Delta\lambda$=42.5\AA\ is the width of that bandpass, and the
pseudo-continuum is interpolated from the \mgtwo\ continuum bands.

In fact we will often find it convenient to express the \mgb\ index in
magnitudes rather than as an equivalent width. By analogy with the
\mgtwo\ index, we therefore define \mgbp\ to be
\begin{equation}
\mgbp = -2.5\log_{10}\left(1-\frac{\mgb}{\Delta\lambda}\right) ~,
\label{eqn:mgbprime}
\end{equation}
where in this case $\Delta\lambda$=32.5\AA, the width of the \mgb\
feature bandpass.

In passing it should be noted that a different definition of
linestrength indices has sometimes been used (\eg\ Worthey 1994,
equations~4 and~5) in which the integral of the ratio of the object
spectrum and the continuum in equations~\ref{eqn:mgbdef}
and~\ref{eqn:mg2def} is replaced by the ratio of the integrals. This
alternative definition has merits (such as simplifying the error
properties of measured indices), but it is not mathematically equivalent
to the standard definition. In practice, however, the two definitions
generally give linestrengths with negligibly different numerical values.

It is usual in studies of this sort to employ the \mgtwo\ index as the
main indicator of metallicity and star-formation history. However we
find it useful for operational reasons to also measure the \mgb\
index. One problem is that the limited wavelength coverage of the
spectra from some runs means that in a number of cases we cannot measure
the \mgtwo\ index (requiring as it does a wider wavelength range)
although we can measure the \mgb\ index. We obtain \mgb\ for 676 objects
(with 299 having repeat measurements) and \mgtwo\ for 582 objects (with
206 having repeat measurements). Another problem with \mgtwo\ is that
the widely-separated continuum bands make it more susceptible than \mgb\
to variations in the non-linear continuum shape of our unfluxed spectra,
which result from using a variety of different instruments and observing
galaxies over a wide range in redshift. We therefore present
measurements of both \mgb\ and \mgtwo: the former because it is
better-determined and available for more sample galaxies, the latter for
comparison with previous work. As previously demonstrated (Gorgas \etal\
1990, J{\o}rgensen 1997) and confirmed here, \mgb\ and \mgtwo\ are
strongly correlated, and so can to some extent be used interchangeably.

Several corrections must be applied to obtain a linestrength measurement
that is calibrated to the standard Lick system. The first correction
allows for the fact that the measured linestrength depends on the
instrumental resolution. Since all our spectra were obtained at higher
resolution than the spectra on which the Lick system was defined, we
simply convolve our spectra with a Gaussian of dispersion
$(\sigma_{Lick}^2-\sigma_i^2)^{1/2}$ in order to broaden our
instrumental resolution $\sigma_i$ (see Table~\ref{tab:obsruns}) to the
Lick resolution of 200\kms.

The second correction allows for the fact that the measured linestrength
depends on the galaxy's internal velocity dispersion---a galaxy with
high enough velocity dispersion $\sigma_g$ will have features broadened
to the point that they extend outside their index bandpasses, and so
their linestrengths will be underestimated.  Moreover, if an absorption
feature is broadened into the neighbouring continuum bands then the
estimated continuum will be depressed and the linestrength will be
further reduced. The `$\sigma$-correction' needed to calibrate out this
effect can be obtained either by measuring linestrength as a function of
velocity broadening for a set of suitable stellar spectra (such as the
templates obtained for measuring redshifts and dispersions) or by
modelling the feature in question.

Although most previous studies have adopted the former approach, we
prefer to use a model to calibrate our indices, since we observe a
dependence of the \mgb\ profile shape on $\sigma$ that is not taken into
account by simply broadening stellar templates. Our simple model assumes
\mgb\ to be composed of three Gaussians centred on the three Mg~I lines
at $\lambda_b$=5166.6\AA, $\lambda_c$=5172.0\AA\ and
$\lambda_r$=5183.2\AA\ with corresponding relative strengths varying
linearly with dispersion from 1.0:1.0:1.0 at $\sigma$=100\kms\ to
0.2:1.0:0.7 at $\sigma$=300\kms. This dependence on dispersion is
empirically determined and approximate (the relative strengths of the
individual lines are not tightly constrained), but it does significantly
improve the profile fits compared to assuming any fixed set of relative
weights. Such variation of the \mgb\ profile shape reflects changes, as
a function of velocity dispersion, in the stellar population mix and
relative abundances (particularly of Mg, C, Fe and Cr), which each
affect the profile in complex ways (Tripicco \& Bell 1995).

Using the estimated value of the index to normalise the model profile
and the effective dispersion $(\sigma_g^2+\sigma_{Lick}^2)^{1/2}$ to
give the broadening, we can estimate both the profile flux which is
broadened out of the feature bandpass and the resulting depression of
the continuum. Correcting for both these effects gives an improved
estimate for the linestrength.  Iterating leads rapidly to convergence
and an accurate $\sigma$-correction for the \mgb\ and \mgtwo\
indices. We find that the \mgb\ $\sigma$-correction is typically +4\% at
100\kms\ and increases approximately linearly to +16\% at 400\kms; the
\mgtwo\ $\sigma$-correction is typically 0.000~mag up to 200\kms\ and
increases approximately linearly to 0.004~mag at 400\kms.

Note that the usual method of determining the $\sigma$-correction by
broadening standard stars ignores the dependence of profile shape on
changes in the stellar population mix as a function of luminosity or
velocity dispersion. Our tests indicate that by doing so, the usual
method tends to overestimate \mgb\ for galaxies with large dispersions:
by 2\% at 200\kms, 6\% at 300\kms\ and 14\% at 400\kms. The two methods
give essentially identical results for \mgtwo, since it has much smaller
$\sigma$-corrections due to its wider feature bandpass and
well-separated continuum bands.

The other corrections that need to be applied to the linestrength
estimates are: (i)~an aperture correction to account for different
effective apertures sampling different parts of the galaxy
(\S\ref{ssec:apcorr}); (ii)~a run correction to remove systematic scale
errors between different observing setups (\S\ref{ssec:combruns}); and
(iii)~an overall calibration to the Lick system determined by
comparisons with literature data (\S\ref{ssec:compare}).

\subsection{Error Estimates}
\label{ssec:errors}

Error estimates for our redshifts, velocity dispersions and
linestrengths come from detailed Monte Carlo simulations of the
measurement process for each observing run. By calibrating the errors
estimated from these simulations against the rms errors obtained from
the repeat measurements that are available for many of the objects (see
\S\ref{ssec:caliberr}), we can obtain precise and reliable error
estimates for each measurement of every object in our sample.

The procedure for estimating the uncertainties in our redshifts and
velocity dispersions was as follows. For each stellar template in each
observing run, we constructed a grid of simulated spectra with Gaussian
broadenings of 100--300\kms\ in 20\kms\ steps and continuum counts
corresponding to $S/N$ ratios of 10--90 in steps of 10. For each
spectrum in this grid we generated 16 realisations assuming Poisson
noise. These simulated spectra were then cross-correlated against all
the other templates from the run in order to derive redshifts and
velocity dispersions in the standard manner. The simulations do not
account for spectral mismatch between the galaxy spectra and the stellar
templates, but for well-chosen templates this effect is only significant
at higher $S/N$ than is typically found in our data.

\begin{figure*}
\plotfull{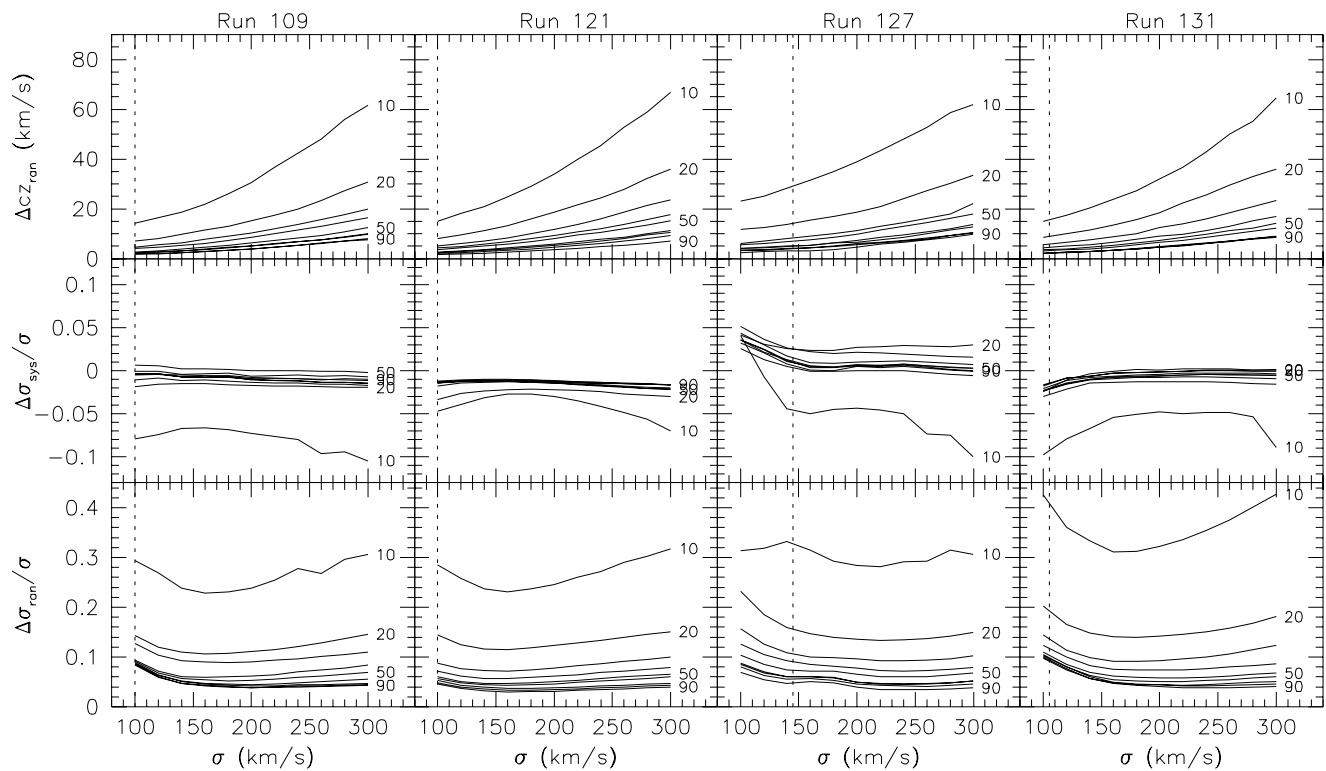}
\caption{Redshift and dispersion errors as functions of input dispersion
and $S/N$ (labelling the curves) from the simulations of four of the
larger runs. The top panel shows the random error in the redshift and
the centre and bottom panels show the systematic and random error in the
dispersion. The vertical dotted line indicates the instrumental
dispersion of each run.
\label{fig:simerrs}}
\end{figure*}

Figure~\ref{fig:simerrs} shows the random error in redshift and the
systematic and random errors in dispersion as functions of input
dispersion and $S/N$ for four of the larger runs. The systematic errors
in redshift are not shown as they are negligibly small ($\sim$1\kms),
although the simulations do not include possible zeropoint errors. The
systematic errors in dispersion are generally small (a few percent or
less) for $S/N$$>$20, but become rapidly larger at lower $S/N$. The
random errors in redshift increase for lower $S/N$ and higher
dispersion, while the random errors in dispersion increase for lower
$S/N$ but have a broad minimum at around twice the instrumental
dispersion. These curves have the general form predicted for the random
errors from the cross-correlation method (Tonry \& Davis 1979, Colless
1987).

Given the dispersion and $S/N$ measured for a spectrum, we interpolated
the error estimates from the simulation for that particular observing
run to obtain the systematic and random errors in each measured
quantity. We used the results of these simulations to correct the
systematic errors in the velocity dispersions and to estimate the
uncertainties in individual measurements of redshift and dispersion. For
quality class D measurements of redshifts, where the spectra are too
poor to estimate a dispersion and hence a reliable redshift error, we
take a conservative redshift error of 50\kms.

The linestrength error estimates were obtained by generating 50 Monte
Carlo realizations of the object spectrum with Poisson noise appropriate
to the spectrum's $S/N$ level. The \mgb\ and \mgtwo\ linestrengths were
then measured for each of these realizations and the error estimated as
the rms error of these measurements about the observed value. The error
estimate obtained in this fashion thus takes into account the noise
level of the spectrum, but does not account for errors in the
linestrength due to errors in the redshift and dispersion estimates, nor
for systematic run-to-run differences in the underlying continuum shape.

The estimated errors in the spectroscopic parameters are compared with,
and calibrated to, the rms errors derived from repeat observations in
\S\ref{ssec:caliberr}.

\subsection{Aperture Corrections}
\label{ssec:apcorr}

The velocity dispersion measured for a galaxy is the luminosity-weighted
velocity dispersion integrated over the region of the galaxy covered by
the spectrograph aperture. It therefore depends on (i)~the velocity
dispersion profile; (ii)~the luminosity profile; (iii)~the distance of
the galaxy; (iv)~the size and shape of the spectrograph aperture; and
(v)~the seeing in which the observations were made. In order to
intercompare dispersion measurements it is therefore necessary to
convert them to a standard scale. The `aperture correction' this
requires has often been neglected because it depends in a complex manner
on a variety of quantities some of which are poorly known. The neglect
of such corrections may account in part for the difficulties often found
in reconciling dispersion measurements from different sources.

The aperture correction applied by Davies \etal\ (1987) was derived by
measuring dispersions for a set of nearby galaxies through apertures of
4\arcsec$\times$4\arcsec and 16\arcsec$\times$16\arcsec.  In this way
they used their nearby galaxies to define the velocity dispersion
profile and obtained a relation between the corrected value,
$\sigma_{cor}$, and the observed one, $\sigma_{obs}$.  This turned out
to be an approximately linear relation amounting to a 5\% correction
over the distance range between Virgo and Coma.

More recently J{\o}rgensen \etal\ (1995) have derived an aperture
correction from kinematic models based on data in the literature.
Published photometry and kinematics for 51 galaxies were used to
construct two-dimensional models of the surface brightness, velocity
dispersion, and rotational velocity projected on the sky. They found
that the position angle only gave rise to 0.5\% variations in the
derived dispersions and could thus be ignored. They converted
rectangular apertures into an `equivalent circular aperture' of radius
$r_{ap}$ which the models predicted would give the same dispersion as
the rectangular slit. They found that to an accuracy of 4\% one could
take $r_{ap} = 1.025(xy/\pi)^{1/2}$, where $x$ and $y$ are the width and
length of the slit.

\begin{figure}
\plotone{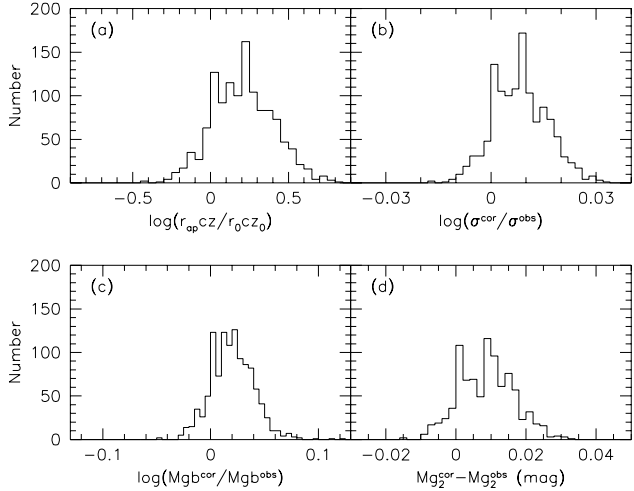}
\caption{The distribution of aperture corrections to a standard metric
aperture. (a)~The distribution of the ratio of the observed metric
apertures to the standard metric aperture (corresponding to 1.7~arcsec
at the redshift of Coma). (b)-(d)~The aperture corrections to this
standard metric aperture for the dispersion, \mgb\ and \mgtwo\
measurements. Note that $\sigma^{cor}$ is the aperture-corrected
dispersion and $\sigma^{obs}$ is the raw observed dispersion; likewise
for the linestrengths.
\label{fig:apcor}}
\end{figure}

From their models they then calculated the correction factor from the
observed dispersion to the dispersion in some standard aperture.  For a
standard {\em metric} aperture, they found this aperture correction to
be well approximated by a power law of the form
\begin{equation}
\frac{\sigma_{cor}}{\sigma_{obs}} = 
     \left[ \left(\frac{r_{ap}}{r_0}\right)
            \left(\frac{cz}{cz_0}\right) \right]^{0.04} ~,
\label{eqn:apcor1}
\end{equation}
where $\sigma_{obs}$ and $\sigma_{cor}$ are the observed and corrected
dispersions, $r_0$ is a standard aperture radius, defined to be
1.7~arcsec, and $cz_0$ is a standard redshift, defined as the redshift
of Coma. The standard metric aperture is thus 0.54\kpc\ in radius.
Alternatively, one can correct to a standard {\em relative} aperture
(defined to be $R_e/8$) using the same power law relation,
\begin{equation}
\frac{\sigma_{cor}}{\sigma_{obs}}=\left(\frac{r_{ap}}{R_e/8}\right)^{0.04} ~.
\label{eqn:apcor2}
\end{equation}
This power law approximates the true relation to within 1\% over the
observed range of effective apertures (compare the distribution metric
aperture sizes in Figure~\ref{fig:apcor}a with Figure~4c of J{\o}rgensen
\etal).

We also apply an aperture correction to our linestrengths. J{\o}rgensen
\etal\ noted that the radial gradient in the \mgtwo\ index is similar to
the radial gradient in $\log\sigma$, and so applied the same aperture
correction for \mgtwo\ as for $\log\sigma$. We adopt this procedure for
\mgtwo. For \mgb\ we convert to \mgbp\ (Equation~\ref{eqn:mgbprime})
and, assuming that the radial profile of \mgbp\ is similar to that of
\mgtwo\ (and hence $\log\sigma$), we apply the $\log\sigma$ aperture
correction to \mgbp\ before converting back to \mgb.

The distributions of corrections to the standard metric aperture for the
dispersions and linestrengths are shown in Figures~\ref{fig:apcor}b-d.
These corrections are generally positive, as most objects in our sample
are observed through larger effective apertures and are further away
than J{\o}rgensen \etal's standard aperture and redshift. The
corrections to standard relative apertures are quite similar, although
having slightly greater amplitude and range. We choose to adopt the
correction to a standard metric aperture in order to minimise the size
and range of the corrections and to facilitate comparisons with
dispersions and linestrengths in the literature.

\subsection{Combining Different Runs}
\label{ssec:combruns}

In comparing the redshifts, dispersions and linestrengths obtained from
different runs we found some significant systematic offsets. The origin
of these run-to-run offsets is not fully understood. For the redshifts,
the use of different velocity standard stars as the fiducials in
different runs clearly contributes some systematic errors. For the
dispersions, the calibration procedure we use should in principle remove
instrumental systematics; in practice, scale differences are common, as
is shown by the range of scale factors needed to reconcile velocity
dispersions from various sources in the compilation by McElroy (1995;
see Table~2).

\begin{table*}
\centering
\caption{Calibration of observing runs to a common system.}
\label{tab:runcorr}
\begin{tabular}{llrrlrrlrrlrr}
Run   & \multicolumn{1}{c}{$\Delta cz$}        & $N_{z}$      & $N^c_{z}$      & 
        \multicolumn{1}{c}{$\Delta\log\sigma$} & $N_{\sigma}$ &	$N^c_{\sigma}$ & 
        \multicolumn{1}{c}{$\Delta$\mgbp}      & $N_{b}$      &	$N^c_{b}$      & 
        \multicolumn{1}{c}{$\Delta$\mgtwo}     & $N_{2}$      & $N^c_{2}$      \\
      & \multicolumn{1}{c}{(\kms)} & & &
        \multicolumn{1}{c}{(dex)}  & & &
        \multicolumn{1}{c}{(mag)}  & & &
        \multicolumn{1}{c}{(mag)}  & & \vspace*{6pt} \\
101     & \n\n$-$9 $\pm$ \n9 &  19 &  62 & $+$0.014 $\pm$ 0.016 &  18 &  65 & $+$0.009 $\pm$ 0.009 &  18 &  32 & $-$0.008 $\pm$ 0.007 &  16 &  30 \\
102     &  \n$+$62 $\pm$  10 &  56 &  57 & $+$0.034 $\pm$ 0.023 &  56 &  56 & $+$0.012 $\pm$ 0.011 &  56 &  24 & $-$0.013 $\pm$ 0.010 &  56 &  12 \\
103     &  \n$-$43 $\pm$  10 &  36 &  53 & $-$0.034 $\pm$ 0.028 &  34 &  48 & $-$0.002 $\pm$ 0.011 &  34 &  27 & $-$0.025 $\pm$ 0.009 &  34 &  20 \\
104$^b$ & \n\n$+$0           &  11 &   0 & $+$0.000             &  11 &   0 & $+$0.000             &  11 &   0 & $+$0.000             &  11 &   0 \\
105     &  \n$+$16 $\pm$ \n9 &  36 &  54 & $+$0.018 $\pm$ 0.019 &  36 &  55 & $+$0.020 $\pm$ 0.007 &  36 &  37 & $+$0.008 $\pm$ 0.009 &  21 &  14 \\
106     &  \n$-$17 $\pm$  14 &  23 &  61 & $+$0.024 $\pm$ 0.029 &  22 &  54 & $+$0.009 $\pm$ 0.007 &  22 &  41 & $+$0.001 $\pm$ 0.014 &  14 &  13 \\
107     &  \n$-$19 $\pm$  10 &  27 &  61 & $+$0.060 $\pm$ 0.033 &  27 &  64 & $+$0.000 $\pm$ 0.009 &  27 &  33 & $-$0.003 $\pm$ 0.008 &  21 &  24 \\
108     &  \n$-$71 $\pm$  23 &   9 &  35 & $-$0.055 $\pm$ 0.035 &   9 &  34 & $-$0.005 $\pm$ 0.008 &   9 &  12 & $+$0.032 $\pm$ 0.008 &   9 &  11 \\
109$^a$ & \n\n$+$1 $\pm$ \n4 &  93 & 222 & $-$0.015 $\pm$ 0.005 &  92 & 220 & $+$0.000 $\pm$ 0.000 &  92 & 167 & $+$0.008 $\pm$ 0.002 &  92 & 126 \\
110     & \n\n$+$3 $\pm$ \n6 &  71 & 186 & $-$0.010 $\pm$ 0.008 &  61 & 171 & $+$0.004 $\pm$ 0.004 &  61 &  78 & $-$0.009 $\pm$ 0.004 &  61 &  50 \\
111     &  \n$-$10 $\pm$  14 &  19 &  72 & $-$0.024 $\pm$ 0.024 &  19 &  76 & $-$0.005 $\pm$ 0.006 &  19 &  33 & $+$0.046 $\pm$ 0.006 &  19 &  25 \\
112     &  \n$+$45 $\pm$ \n8 &  31 &  82 & $-$0.006 $\pm$ 0.008 &  31 & 103 & $+$0.001 $\pm$ 0.009 &  31 &  16 & $-$0.027 $\pm$ 0.013 &  16 &   7 \\
113     &   $+$154 $\pm$  15 &  20 &   9 & $+$0.041 $\pm$ 0.038 &  20 &   9 & $+$0.025 $\pm$ 0.012 &  20 &   7 & $-$0.015 $\pm$ 0.026 &   2 &   1 \\
114     & \n\n$-$9 $\pm$  11 &  12 &  22 & $-$0.059 $\pm$ 0.024 &  12 &  22 & $-$0.012 $\pm$ 0.009 &  12 &  20 & $+$0.032 $\pm$ 0.007 &  12 &  15 \\
115     & \n\n$+$9 $\pm$  10 &   8 &  24 & $+$0.024 $\pm$ 0.022 &   8 &  23 & $-$0.069 $\pm$ 0.026 &   1 &   2 & $-$0.087 $\pm$ 0.034 &   1 &   2 \\
116$^d$ &  ~~~~~~~---        & --- & --- & ~~~~~~~~~~---        & --- & --- & ~~~~~~~~~~---        & --- & --- & ~~~~~~~~~~---        & --- & --- \\
117     & \n\n$-$2 $\pm$ \n7 &  59 & 132 & $+$0.005 $\pm$ 0.021 &  55 & 121 & $+$0.016 $\pm$ 0.007 &  55 &  90 & $+$0.028 $\pm$ 0.007 &  41 &  51 \\
118$^c$ &   $+$120 $\pm$  22 &  14 &   4 & $-$0.018 $\pm$ 0.066 &  13 &   4 & $-$0.011 $\pm$ 0.029 &  13 &   3 & $+$0.000             &   5 &   0 \\
119     &  \n$-$20 $\pm$ \n9 &  17 &  20 & $-$0.004 $\pm$ 0.015 &  17 &  19 & $+$0.003 $\pm$ 0.007 &  17 &  19 & $-$0.013 $\pm$ 0.009 &  17 &   8 \\
120     &  \n$-$39 $\pm$ \n7 &  38 &  47 & $+$0.009 $\pm$ 0.011 &  34 &  44 & $+$0.005 $\pm$ 0.005 &  33 &  25 & $+$0.059 $\pm$ 0.008 &  26 &  14 \\
121     &  \n$-$66 $\pm$ \n8 &  86 & 177 & $+$0.038 $\pm$ 0.010 &  82 & 181 & $+$0.002 $\pm$ 0.008 &  82 &  23 & $+$0.008 $\pm$ 0.008 &  54 &  17 \\
122     &  \n$-$28 $\pm$ \n8 &  41 &  70 & $+$0.001 $\pm$ 0.012 &  37 &  58 & $+$0.020 $\pm$ 0.006 &  37 &  32 & $+$0.033 $\pm$ 0.008 &  31 &  24 \\
123     &  \n$-$22 $\pm$ \n8 &  22 &  41 & $+$0.022 $\pm$ 0.017 &  17 &  34 & $+$0.005 $\pm$ 0.006 &  16 &  26 & $+$0.010 $\pm$ 0.006 &  13 &  16 \\
124     &  \n$+$14 $\pm$  17 &  22 &  49 & $+$0.020 $\pm$ 0.044 &  14 &  40 & $+$0.029 $\pm$ 0.012 &  14 &   8 & $+$0.012 $\pm$ 0.022 &  11 &   1 \\
125     &  \n$-$48 $\pm$  14 &  57 &  62 & $+$0.037 $\pm$ 0.010 &  55 &  64 & $+$0.007 $\pm$ 0.008 &  55 &   8 & $-$0.018 $\pm$ 0.011 &  55 &   7 \\
126     &  \n$+$57 $\pm$  14 &  36 &  43 & $-$0.004 $\pm$ 0.039 &  33 &  40 & $+$0.029 $\pm$ 0.023 &  33 &   9 & $+$0.008 $\pm$ 0.023 &  33 &   6 \\
127     & \n\n$-$3 $\pm$ \n4 & 131 & 187 & $+$0.002 $\pm$ 0.007 & 127 & 167 & $-$0.007 $\pm$ 0.003 & 127 & 136 & $-$0.007 $\pm$ 0.003 & 127 &  83 \\
128     & \n\n$+$6 $\pm$  12 &  24 &  29 & $-$0.042 $\pm$ 0.010 &  23 &  29 & $+$0.010 $\pm$ 0.015 &  23 &   7 & $-$0.055 $\pm$ 0.013 &   9 &   4 \\
129$^b$ & \n\n$+$0           &   3 &   0 & $+$0.000             &   3 &   0 & $+$0.000             &   3 &   0 & $+$0.000             &   3 &   0 \\
130$^d$ &  ~~~~~~~---        & --- & --- & ~~~~~~~~~~---        & --- & --- & ~~~~~~~~~~---        & --- & --- & ~~~~~~~~~~---        & --- & --- \\
131$^e$ &  \n$+$35 $\pm$ \n5 & 123 & 174 & $-$0.078 $\pm$ 0.014 &  99 & 152 & $+$0.002 $\pm$ 0.004 &  98 & 136 & ~~~~~~~~~~---        & --- & --- \\
132     &  \n$-$24 $\pm$  19 &  12 &   7 & $-$0.025 $\pm$ 0.026 &  12 &   6 & $-$0.032 $\pm$ 0.008 &  12 &   4 & $+$0.005 $\pm$ 0.016 &  12 &   4 \\
133     &  \n$-$11 $\pm$ \n4 & 128 & 247 & $-$0.009 $\pm$ 0.006 & 128 & 241 & $-$0.009 $\pm$ 0.002 & 128 & 183 & $-$0.014 $\pm$ 0.002 & 128 & 151 \\
\end{tabular}\vspace*{6pt}
\parbox{0.84\textwidth}{
$^a$ Run 109 is the fiducial run for \mgbp, defined to have zero offset. \\ 
$^b$ Runs 104 and 129 have no objects in common with other runs. \\ 
$^c$ Run 118 has no \mgtwo\ measurements in common with other runs. \\
$^d$ Runs 116 and 130 have no usable data. \\
$^e$ Run 131 has no \mgtwo\ measurements. }
\end{table*}

We cannot directly calibrate the measurements from each run to the
system defined by a chosen fiducial run, as there is no run with objects
in common with all other runs to serve as the fiducial. Instead, we use
the mean offset, $\Delta$, between the measurements from any particular
run and {\em all} the other runs. To compute this offset we separately
compute, for each galaxy $i$, the error-weighted mean value of the
measurements obtained from the run in question, $x_{ij}$, and from all
other runs, $y_{ik}$:
\begin{equation}
\langle x_i \rangle = \frac{\sum_j x_{ij}/\delta_{ij}^2}
                           {\sum_j 1/\delta_{ij}^2} ~~,~~ 
\langle y_i \rangle = \frac{\sum_k y_{ik}/\delta_{ik}^2}
                           {\sum_k 1/\delta_{ik}^2} ~.
\label{eqn:meangal}
\end{equation}
Here $j$ runs over the $m_i$ observations of galaxy $i$ in the target
run and $k$ runs over the $n_i$ observations of galaxy $i$ in all other
runs; $\delta_{ij}$ and $\delta_{ik}$ are the estimated errors in
$x_{ij}$ and $y_{ik}$. We then take the average over all galaxies,
weighting by the number of comparison pairs, to arrive at an estimate
for the offset of the target run:
\begin{equation}
\Delta = \frac{\sum_i m_i n_i (\langle x_i \rangle - \langle y_i \rangle)}
              {\sum_i m_i n_i}
\label{eqn:offset}
\end{equation}
Here $i$ runs over the $l$ galaxies in the sample. We can reject
outliers at this point by excluding galaxies for which the difference
$\langle x_i \rangle - \langle y_i \rangle$ is larger than some cutoff:
for $cz$, $\log\sigma$, \mgbp\ and \mgtwo\ we required differences less
than 300\kms, 0.2~dex, 0.1~mag and 0.1~mag respectively. The
uncertainty, $\epsilon$, in this estimate of the run offset is given by
\begin{equation}
\epsilon^2 = \frac{\sum_i (m_i n_i)^2 
  (\delta\langle x_i \rangle^2+\delta\langle y_i \rangle^2)}
  {\bigl(\sum_i m_i n_i\bigr)^2}
\label{eqn:offerr}
\end{equation}
where $\delta\langle x_i \rangle$ and $\delta\langle y_i \rangle$ are
the error-weighted uncertainties in $\langle x_i \rangle$ and 
$\langle y_i \rangle$ given by
\begin{equation}
\textstyle
\delta\langle x_i \rangle^2 = \bigl(\sum_j\delta_{ij}^{-2}\bigr)^{-1} ~~,~~
\delta\langle y_i \rangle^2 = \bigl(\sum_k\delta_{ik}^{-2}\bigr)^{-1} ~.
\label{eqn:galerr}
\end{equation}
We subtract the offset determined in this manner from each run and then
iterate the whole procedure until there are no runs with residual
offsets larger than $0.5\epsilon$. As a final step, we place the entire
dataset (now corrected to a common zeropoint) onto a fiducial system by
subtracting from all runs the offset of the fiducial system. Note that
the run corrections for dispersion and \mgb\ are determined in terms of
offsets in $\log\sigma$ and \mgbp.

In order to maximise the number of objects with multiple measurements,
we included the dataset from the `Streaming Motions of Abell Clusters'
project (SMAC: M.J.Hudson, priv.comm.; see also Smith \etal\ 1997) in
this analysis. There is a considerable overlap between the SMAC and EFAR
samples which significantly increases the number of comparison
observations and reduces the uncertainties in the run offsets. We chose
to use the `Lick' system of Davies \etal\ (1987; included in the SMAC
dataset) as our fiducial, in order to bring the 7~Samurai, EFAR and SMAC
datasets onto a single common system. This is not possible with \mgb,
which is not measured in most previous work or by SMAC. We therefore
chose run~109 (the Kitt Peak 4m run of November 1988) as the \mgb\
fiducial because it had a large number of high-quality observations and
the systematics of the slit spectrograph are believed to be well
understood.

We checked that this procedure gives relative run corrections consistent
with those obtained by directly comparing runs in those cases where
there {\em are} sufficient objects in common. We have also compared our
method with a slightly different method used by the SMAC collaboration
to determine the run corrections for their own data and found good
agreement (M.J.Hudson, priv.comm.). We carried out Monte Carlo
simulations of the whole procedure in order to check the uncertainties
in the offsets computed according to Equation~\ref{eqn:offerr}. We found
that this equation in general provides a good estimate of the
uncertainties, although when the number of comparisons is small or
involve a small number of other runs it can under-estimate the
uncertainties by up to 30\%. Our final estimates of the uncertainties
are therefore derived as the rms of the offsets from 100 Monte Carlo
simulations.

Table~\ref{tab:runcorr} lists the offsets for each run computed
according to the above procedure, their uncertainties based on Monte
Carlo simulations, the number of individual measurements ($N$) and the
number of comparison pairs ($N^c$). Note that to correct our observed
measurements to the fiducial system we {\em subtract} the appropriate
run offset in Table~\ref{tab:runcorr} from each individual measurement.
Of the 31 spectroscopic runs with usable data, only runs 104 and 129
have no objects in common with other runs and hence no run corrections;
run 118 has no \mgtwo\ measurements in common and so no run correction
for \mgtwo.

Weighting by the number of individual measurements in each run, the mean
amplitude of the corrections and their uncertainties are 28$\pm$8\kms\
in $cz$, 0.023$\pm$0.015~dex in $\log\sigma$, 0.008$\pm$0.006~mag in
\mgbp\ and 0.015$\pm$0.006~mag in \mgtwo. The significance of the
individual run corrections (in terms of the ratio of the amplitude of
the offset to its uncertainty) varies; however over all runs the reduced
$\chi^2$ is highly significant: 15.7, 4.0, 3.3 and 11.4 for the
corrections to the redshifts, dispersions, \mgb\ and \mgtwo\
respectively. Application of the run corrections reduces the median rms
error amongst those objects with repeat measurements from 18\kms\ to
14\kms\ in redshift, 6.3\% to 5.6\% in dispersion, 4.9\% to 4.4\% in
\mgb\ and 0.012~mag to 0.009~mag in \mgtwo. We also checked to see
whether applying the run corrections reduced the scatter in external
comparisons between our data and measurements in the literature (see
\ref{ssec:compare}). We found that although the scatter is dominated by
the combined random errors, the corrections did reduce the scatter
slightly in all cases.

As another test of the run corrections for \mgbp\ and \mgtwo\ (and also,
more weakly, for $\log\sigma$), we compared the \mgbpsig\ and \mgtwosig\
distributions for each run (after applying the run corrections) with the
global \mgbpsig\ and \mgtwosig\ relations derived in Paper~V. Using the
$\chi^2$ goodness-of-fit statistic to account both for measurement
errors in the dispersions and linestrengths and for the intrinsic
scatter about the \mgsig\ relations, we find that for \mgbpsig\ there
were two runs (113 and 132) with reduced $\chi^2$ greater than 3, while
for \mgtwosig\ there was one such run (122). In all three cases the
removal of 1 or 2 obvious outliers decreased the reduced $\chi^2$ to a
non-significant level.

\subsection{Calibrating the Estimated Errors}
\label{ssec:caliberr}

\begin{figure*}
\plotfull{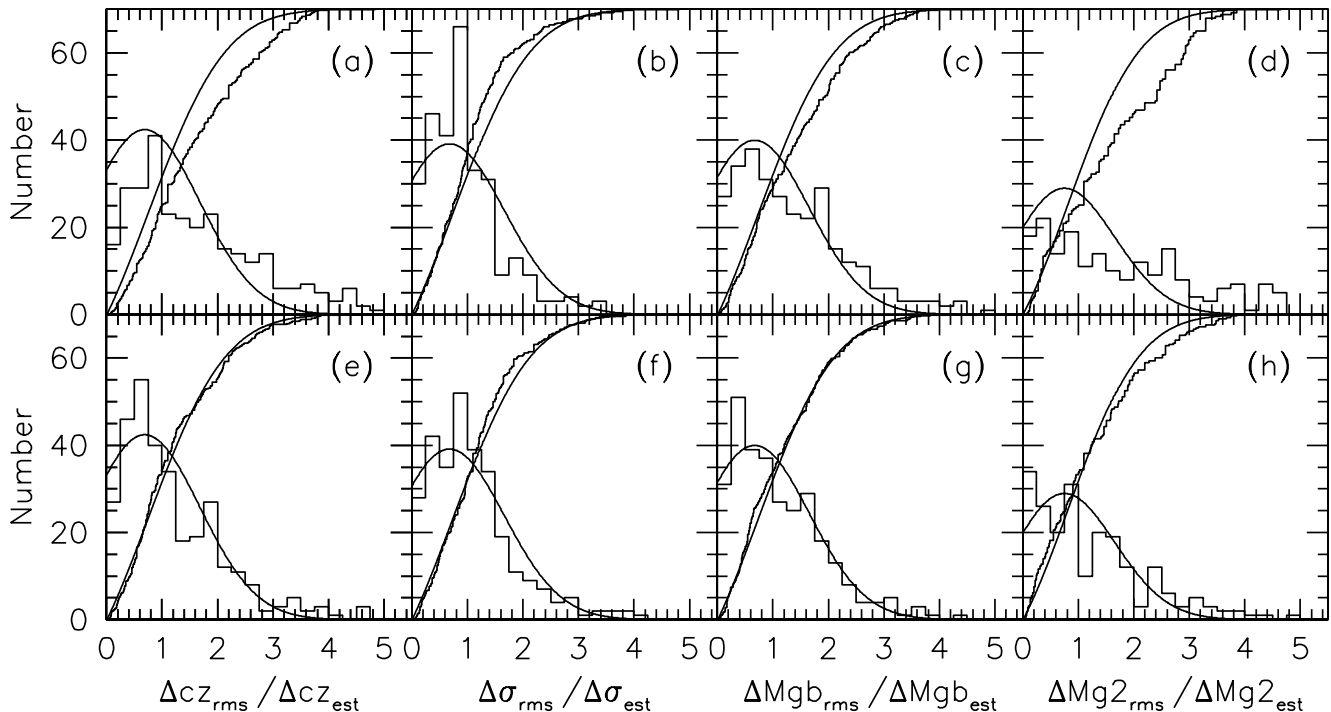}
\caption{Comparison of the estimated errors derived from simulations and
the rms errors for galaxies with repeat measurements of redshift,
dispersion, \mgb\ and \mgtwo. Each panel shows the differential and
cumulative distributions of the ratio of rms error to estimated error.
The stepped curves are the observed distributions, while the smooth
curves are the predicted distributions. The upper panels show the
comparisons using the original estimated errors; the lower panels show
the comparisons after correcting the estimated errors as described in
the text.
\label{fig:errcomp}}
\end{figure*}

Obtaining precise error estimates is particularly important because we
will make extensive use of them in applying maximum likelihood methods
to deriving the Fundamental Plane and relative cluster distances for our
sample. Although we have estimated the measurement errors as carefully
as possible, simulating the noise in the observations and the
measurement procedures, some sources of error are likely to remain
unaccounted-for and we may be systematically mis-estimating the errors.
We therefore auto-calibrate our errors by scaling the estimated errors
in the combined measurements (the {\em internal} error estimate, based
on the individual measurement errors derived from simulations; see
\S\ref{ssec:errors} and \S\ref{ssec:combmeas}) to match the rms errors
from objects with repeat measurements (an {\em external} error
estimate).

Figure~\ref{fig:errcomp} shows the differential and cumulative
distributions of the ratio of the rms error to the estimated error for
each galaxy with repeat measurements of redshift, dispersion, \mgb\ and
\mgtwo. The smooth curves are the predicted differential and cumulative
distributions of this ratio assuming that the estimated errors are the
true errors. The top panel shows the comparison using the estimated
errors (including all the corrections discussed above). For the
redshifts and linestrengths, the estimated errors are generally
under-estimates of the true errors, since the ratio of rms to estimated
errors tends to be larger than predicted. For the dispersions the
estimated errors are generally over-estimates of the true errors, since
this ratio tends to be smaller than predicted. For all quantities the
assumption that the estimated errors are consistent with the true errors
is ruled out with high confidence by a Kolmorogorov-Smirnov (KS) test
applied to the observed and predicted cumulative distributions. These
differences between the estimated errors from the simulations and the
rms errors from repeat measurements reflect the approximate nature of
the $S/N$ estimates and systematic measurement errors not accounted for
in the simulations.

In order to bring our estimated errors into line with the rms errors
from the repeat measurements, we found it necessary to add 15\kms\ in
quadrature to the estimated redshift errors, to scale the dispersion and
\mgb\ errors by factors of 0.85 and 1.15 respectively, and to add
0.005~mag to the \mgtwo\ errors. These corrections were determined by
maximising the agreement of the observed and predicted distributions of
the ratio of rms to estimated errors under a KS test (excluding outliers
with values of this ratio $>$3.5). The corrections are quite well
determined: to within a couple of \kms\ for the redshift correction, a
few percent for the dispersion and \mgb\ corrections, and 0.001~mag for
the \mgtwo\ correction. Applying these corrections and repeating the
comparison of rms and estimated errors gives the lower panels of
Figure~\ref{fig:errcomp}, which shows the good agreement between the rms
errors from repeat measurements and the calibrated errors estimates for
the redshifts, dispersions and Mg linestrengths.

The need for such a correction to the redshift errors may be due in part
to the residual zeropoint uncertainties in the redshifts and in part to
a tendency for the simulations to under-estimate the errors for high
$S/N$ spectra. The origin of the over-estimation of the dispersion
errors is uncertain, although it may result from slightly different
prescriptions for estimating the $S/N$ in the observations and the
simulations. The under-estimation of the linestrength errors may be due
to neglecting the effects of errors in the redshift and dispersion
estimates and the different continuum shapes of spectra from different
runs when measuring linestrengths.

\section{RESULTS}
\label{sec:results}

\subsection{Individual Measurements}
\label{ssec:measure}

The previous two sections describe the observations and analysis of our
spectroscopic data. Table~\ref{tab:spectab} lists the observational
details for each spectrum and the fully-corrected measurements of
redshift, dispersion, \mgb\ and \mgtwo, together with their calibrated
error estimates. Note that these error estimates are the individual
measurement errors, and must be combined in quadrature with the run
correction uncertainties given in Table~\ref{tab:runcorr} to give the
total error estimate. We list the measurement errors rather than the
total errors because the total errors are not independent, being
correlated for objects in the same run. The version of the table
presented here is abridged; the full table will be available upon
publication from NASA's Astrophysical Data Center (ADC) and from the
Centre de Donn\'{e}es astronomiques de Strasbourg (CDS).

\begin{table*}
\centering
\caption{Individual spectroscopic measurements (abridged)}
\label{tab:spectab} 
\begin{tabular}{clcclrrrrrrrrrl}
(1) & (2) & (3) & (4) & (5) & (6) & (7) & (8) & (9) & (10) & (11) & (12) & (13) & (14) & (15) \\
{\scriptsize GINRUNSEQ} & Galaxy & Tele- & Obsvn & $Q$ & $S/N$ & \multicolumn{2}{c}{$cz$} & 
\multicolumn{2}{c}{$\sigma$} & \multicolumn{2}{c}{\mgb} & \multicolumn{2}{c}{\mgtwo} & {Notes} \\
 & Name & scope & Date &  & & \multicolumn{2}{c}{(\kms)} & \multicolumn{2}{c}{(\kms)} & 
\multicolumn{2}{c}{(\AA)} & \multicolumn{2}{c}{(mag)} & \vspace*{6pt} \\
001106215 & A76~A     & MG24 & 880914 & A  & 21.4 & 11296 &  51 & 262 &  50 &  5.20 &  0.54 & 0.353 & 0.020 &  \\
002106221 & A76~B     & MG24 & 880914 & B  & 18.8 & 11317 &  47 & 225 &  47 &  4.76 &  0.59 & 0.326 & 0.019 &  \\
003106218 & A76~C     & MG24 & 880914 & A  & 17.8 & 11973 &  49 & 217 &  48 &  5.17 &  0.68 & 0.392 & 0.021 &  \\
004107095 & A76~D     & MG24 & 881011 & C  & 14.8 & 12184 &  31 & 316 & 103 &  6.18 &  0.83 & 0.326 & 0.028 &  \\
005120504 & A76~E     & IN25 & 901016 & B  & 33.2 & 12241 &  19 & 217 &  16 &  6.35 &  0.22 & 0.312 & 0.011 &  \\
005120505 & A76~E     & IN25 & 901017 & B  & 32.8 & 12147 &  17 & 169 &  11 &  6.16 &  0.23 & 0.275 & 0.012 &  \\
006107098 & A76~F     & MG24 & 881011 & B  & 17.6 & 12371 &  28 & 307 &  82 &  5.29 &  0.83 & 0.331 & 0.027 &  \\
008122534 & A85~A     & MG24 & 911016 & B  & 18.7 & 16577 &  39 & 290 &  52 &  5.14 &  0.59 & 0.344 & 0.021 &  \\
008123319 & A85~A     & IN25 & 911201 & D  & 21.4 & 16692 &  50 & --- & --- &   --- &   --- &   --- &   --- &  \\
009120626 & A85~B     & IN25 & 901017 & A  & 37.6 & 17349 &  34 & 436 &  46 &  5.74 &  0.22 &   --- &   --- &  \\
010120628 & A85~C     & IN25 & 901017 & D  & 33.9 & 22837 &  50 & --- & --- &   --- &   --- &   --- &   --- & \mgb\ at sky \\
011122540 & A85~1     & MG24 & 911016 & B  & 25.8 & 15112 &  20 & 195 &  18 &  3.82 &  0.39 & 0.253 & 0.016 &  \\
012132001 & A85~2     & SS2B & 931022 & B  & 28.3 & 16264 &  25 & 294 &  32 &  5.86 &  0.31 & 0.337 & 0.011 &  \\
013101059 & A119~A    & MG24 & 861202 & B  & 16.5 & 11516 &  49 & 299 &  48 &  6.06 &  0.51 & 0.294 & 0.022 &  \\
013109339 & A119~A    & KP4M & 881107 & A  & 44.2 & 11457 &  19 & 289 &  18 &  5.00 &  0.20 & 0.320 & 0.010 &  \\
013131330 & A119~A    & ES36 & 931008 & A  & 30.9 & 11451 &  24 & 320 &  29 &  4.86 &  0.26 &   --- &   --- &  \\
014101063 & A119~B    & MG24 & 861202 & C  & 14.4 & 13205 &  63 & 323 &  65 &  5.90 &  0.72 & 0.320 & 0.025 &  \\
014109339 & A119~B    & KP4M & 881107 & A  & 35.0 & 13345 &  21 & 276 &  21 &  4.86 &  0.22 & 0.361 & 0.010 &  \\
015109343 & A119~C    & KP4M & 881107 & A  & 33.2 & 13508 &  19 & 250 &  19 &  5.51 &  0.22 & 0.295 & 0.011 &  \\
015131330 & A119~C    & ES36 & 931008 & B  & 19.5 & 13484 &  28 & 265 &  35 &  5.88 &  0.48 &   --- &   --- &  \\
016109346 & A119~D    & KP4M & 881107 & C* & 20.8 & 14980 &  16 & 104 &  13 &  3.30 &  0.37 & 0.151 & 0.016 & H$\beta$ \\
016110611 & A119~D    & KP2M & 881114 & D* & 15.6 & 15022 &  50 & --- & --- &   --- &   --- &   --- &   --- & H$\beta$ \\
016131330 & A119~D    & ES36 & 931008 & D* & 16.1 & 14996 &  50 & --- & --- &   --- &   --- &   --- &   --- & H$\beta$ \\
017109347 & A119~E    & KP4M & 881107 & A  & 36.1 & 12807 &  19 & 251 &  17 &  5.18 &  0.18 & 0.326 & 0.010 &  \\
017131330 & A119~E    & ES36 & 931008 & A  & 27.0 & 12788 &  21 & 243 &  22 &  6.01 &  0.32 &   --- &   --- &  \\
018109342 & A119~F    & KP4M & 881107 & B  & 28.4 & 13034 &  18 & 193 &  15 &  4.95 &  0.32 & 0.245 & 0.013 &  \\
018131330 & A119~F    & ES36 & 931008 & C  & 21.1 & 13006 &  16 & 112 &  20 &  4.69 &  0.37 &   --- &   --- &  \\
019109342 & A119~G    & KP4M & 881107 & E  & 32.7 &   --- & --- & --- & --- &   --- &   --- &   --- &   --- & mis-ID \\
019131330 & A119~G    & ES36 & 931008 & A  & 36.7 & 13457 &  19 & 267 &  19 &  4.78 &  0.20 &   --- &   --- &  \\
021122654 & A119~I    & MG24 & 911018 & A  & 31.0 & 13271 &  20 & 225 &  19 &  4.85 &  0.38 & 0.326 & 0.014 &  \\
022131330 & A119~J    & ES36 & 931008 & B  & 22.3 & 13520 &  21 & 219 &  24 &  4.55 &  0.33 &   --- &   --- &  \\
023131330 & A119~1    & ES36 & 931008 & C* & 11.5 &  4127 &  18 &  92 &  38 &  0.50 &  0.78 &   --- &   --- & H$\beta$ \\
024122657 & A119~2    & MG24 & 911018 & E  & 28.9 &   --- & --- & --- & --- &   --- &   --- &   --- &   --- & mis-ID \\
024131330 & A119~2    & ES36 & 931008 & B  & 20.4 & 12346 &  16 & 100 &  20 &  5.13 &  0.38 &   --- &   --- &  \\
025107103 & J3~A      & MG24 & 881011 & B  & 15.2 & 14453 &  32 & 333 & 111 &  6.25 &  0.74 & 0.329 & 0.025 &  \\
026120714 & J3~B      & IN25 & 901018 & A  & 39.5 & 13520 &  18 & 231 &  15 &  4.93 &  0.23 &   --- &   --- &  \\
027107106 & J3~C      & MG24 & 881011 & C  & 15.8 & 13771 &  21 & 163 &  38 &  4.33 &  0.86 & 0.224 & 0.027 &  \\
028120712 & J3~D      & IN25 & 901017 & A  & 36.3 & 14316 &  22 & 287 &  23 &  5.45 &  0.24 &   --- &   --- & double \\
028120713 & J3~D      & IN25 & 901017 & B  & 30.8 & 14837 &  19 & 207 &  15 &  4.87 &  0.26 &   --- &   --- & double \\
031107190 & J4~A      & MG24 & 881012 & C  & 13.2 & 12074 &  24 & 185 &  51 &  5.35 &  0.75 & 0.387 & 0.029 &  \\
032107189 & J4~B      & MG24 & 881012 & C  & 13.3 & 12090 &  31 & 302 & 102 &  4.11 &  1.08 & 0.261 & 0.032 &  \\
033132002 & J4~C      & SS2B & 931021 & A  & 30.0 & 17154 &  30 & 358 &  43 &  6.20 &  0.26 & 0.355 & 0.010 &  \\
036107195 & A147~A    & MG24 & 881012 & C  & 12.2 & 12811 &  26 & 208 &  62 &  3.96 &  1.16 & 0.288 & 0.034 &  \\
036117298 & A147~A    & MG24 & 891015 & B  & 19.0 & 12741 &  35 & 281 &  49 &  4.84 &  0.60 & 0.314 & 0.022 &  \\
036133157 & A147~A    & CT4M & 931019 & A  & 61.4 & 12760 &  17 & 235 &  12 &  4.77 &  0.17 & 0.285 & 0.010 &  \\
036133158 & A147~A    & CT4M & 931019 & A  & 62.3 & 12776 &  17 & 253 &  13 &  5.01 &  0.18 & 0.289 & 0.010 &  \\
\end{tabular}
\vspace*{6pt} \\ \parbox{\textwidth}{This is an abridged version of this
table; the full table will be available upon publication from NASA's
Astrophysical Data Center (ADC) and from the Centre de Donn\'{e}es
astronomiques de Strasbourg (CDS). The columns of this table give:
(1)~observation identifier (GINRUNSEQ); (2)~galaxy name; (3)~telescope
used; (4)~date of observation; (5)~quality parameter; (6)~signal to
noise ratio; (7--8)~redshift and estimated error; (9--10)~velocity
dispersion and estimated error; (11--12)~\mgb\ linestrength and
estimated error; (13--14)~\mgtwo\ linestrength and estimated error; and
(15) notes on each observation. In the notes, `double' means the EFAR
galaxy is double; `star' means the EFAR object is a star not a galaxy;
`mis-ID' means the spectrum is for some galaxy other than the nominated
EFAR object; `mis-ID*' means the spectrum is for a nearby star rather
than the EFAR object; `\mgb\ at sky' means the object is at a redshift
which puts \mgb\ on the 5577\AA\ sky line; `\#=\#' notes the duplicated
pairs in the EFAR sample (see Paper~I; only the first of the two GINs is
used); emission line objects (with an asterisk on $Q$) have the emission
features listed; `H$\beta$ abs' or `H$\beta$ abs, [OIII]' means the
redshift is based on the H$\beta$ absorption feature (and [OIII] if
present), as the spectrum stops short of \mgb\ (no dispersion or \mgb\
index is given for these objects). The objects for which we have no
spectrum have GINs: 7, 20, 29, 30, 34, 35, 55, 62, 64, 67, 82, 83, 91,
104, 121, 131, 133, 134, 161, 181, 191, 214, 225, 228, 231, 234, 256,
265, 309, 327, 391, 405, 407, 417, 434, 435, 442, 450, 451, 452, 458,
463, 464, 465, 470, 475, 477, 483, 484, 486, 494, 516, 520, 521, 522,
523, 526, 544, 551, 553, 567, 569, 570, 575, 576, 577, 587, 594, 597,
603, 605, 624, 625, 644, 671, 727, 760, 793, 798, 801, 901.}
\end{table*}

The entries in Table~\ref{tab:spectab} are as follows: Column~1 gives
GINRUNSEQ, a unique nine-digit identifier for each spectrum, composed of
the galaxy identification number (GIN) as given in the master list of
EFAR sample galaxies (Table~3 of Paper~I), the run number (RUN) as given
in Table~\ref{tab:obsruns}, and a sequence number (SEQ) which uniquely
specifies the observation within the run; column~2 gives the galaxy
name, as in the master list of Paper~I; column~3 is the telescope code,
as in Table~\ref{tab:obsruns}; column~4 is the UT date of the
observation; columns~5 \&~6 are the quality parameter (with an asterisk
if the spectrum shows emission features) and $S/N$ of the spectrum (see
\S\ref{ssec:quality}); columns~7 \&~8 are the fully-corrected
heliocentric redshift $cz$ (in \kms) and its measurement error;
columns~9 \&~10 are the fully-corrected velocity dispersion $\sigma$ (in
\kms) and its measurement error; columns~11 \&~12 are the
fully-corrected \mgb\ linestrength index and its measurement error (in
\AA); columns~13 \&~14 are the fully-corrected \mgtwo\ linestrength
index and its measurement error (in mag); column~15 provides comments,
the full meanings of which are described in the notes to the table.

There are 1319 spectra in this table. Note that 81 objects from our
original sample do not have spectroscopic observations and do not appear
in the table (see the list of missing GINs in the table notes). Three of
these are the duplicate objects (GINs 55, 435, 476) and three are known
stars (GINs 131, 133, 191). Most of the others are objects which our
imaging showed are not early-type galaxies, although there are a few
early-type galaxies for which we did not get a spectrum. There are 34
spectra which are unusable ($Q$=E) either because the spectrum is too
poor (13 cases) or because the object was mis-identified (20 cases) or
is a known star (1 case, GIN 123). Of the 1285 usable spectra (for 706
different galaxies), there are 637 spectra with $Q$=A, 407 with $Q$=B,
161 with $Q$=C and 80 with $Q$=D.

\begin{figure}
\plotone{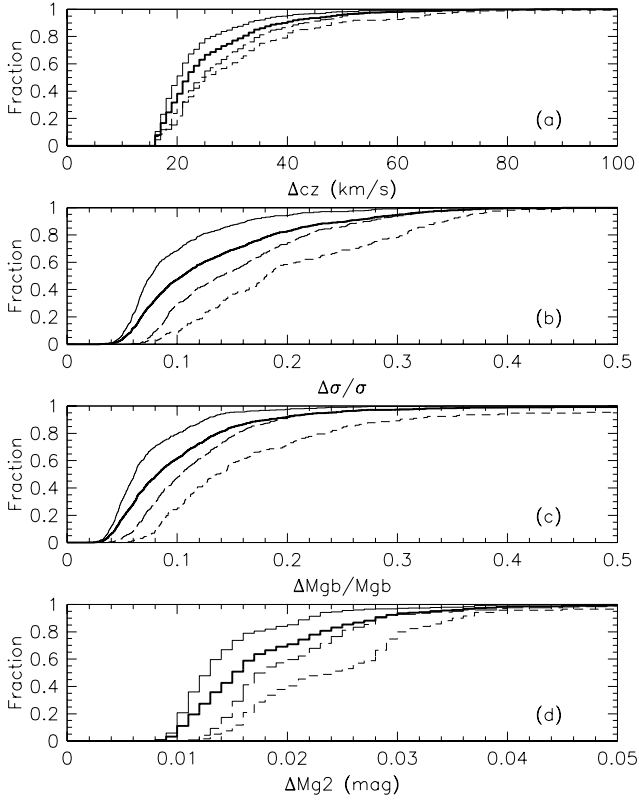}
\caption{The cumulative distributions of estimated errors for individual
measurements of redshift, velocity dispersion and \mgb\ linestrength.
The distributions for quality classes A, B and C are shown as full,
long-dashed and short-dashed lines respectively; the overall
distribution is the thick full line. (a)~The distribution of estimated
errors in $cz$; (b)~estimated relative errors in $\sigma$; (c)~estimated
relative errors in \mgb; (d)~estimated errors in \mgtwo.
\label{fig:errdist1}}
\end{figure}

The cumulative distributions of the total estimated errors in the
individual measurements (combining measurement errors and run correction
uncertainties in quadrature) are shown in Figure~\ref{fig:errdist1} for
quality classes A, B and C, and for all three classes together. The
error distributions can be quantitatively characterised by their 50\%
and 90\% points, which are listed in Table~\ref{tab:typerrs1}. The
overall median error in a single redshift measurement is 22\kms, the
median relative errors in single measurements of dispersion and \mgb\
are 10.5\% and 8.2\%, and the median error in a single measurement of
\mgtwo\ is 0.015~mag.

\subsection{Combining Measurements}
\label{ssec:combmeas}

We use a weighting scheme to combine the individual measurements of each
quantity to obtain a best estimate (and its uncertainty) for each galaxy
in our sample. The weighting has three components:

\begin{table}
\centering
\caption{The distribution of estimated errors per measurement}
\label{tab:typerrs1}
\begin{tabular}{ccccccccc}
$Q$ & \multicolumn{2}{c}{$\Delta cz$ (km/s)} & 
\multicolumn{2}{c}{$\Delta\sigma/\sigma$} & 
\multicolumn{2}{c}{$\Delta$\mgb/\mgb} &
\multicolumn{2}{c}{$\Delta$\mgtwo\ (mag)} \\
    & 50\% & 90\% &  50\% &  90\% &  50\% &  90\% &  50\% &  90\% \\
All & 22   & 40   & 0.105 & 0.255 & 0.082 & 0.184 & 0.015 & 0.028 \\
A   & 20   & 33   & 0.076 & 0.163 & 0.061 & 0.125 & 0.013 & 0.022 \\
B   & 24   & 43   & 0.140 & 0.275 & 0.104 & 0.192 & 0.018 & 0.028 \\
C   & 25   & 48   & 0.181 & 0.343 & 0.136 & 0.303 & 0.024 & 0.036 \\
\end{tabular}
\end{table}

(i) {\em Error weighting:} For multiple measurements $X_i$ having
estimated total errors $\Delta_i$ (the measurement errors and run
correction uncertainties added in quadrature), we weight the values
inversely with their variances, \ie\ by $\Delta_i^{-2}$.

(ii) {\em Quality weighting:} We apply a weighting $W_Q$ which
quantifies our degree of belief (over and above the estimated errors) in
measurements obtained from spectra with different quality parameters.
Following the discussion in \S\ref{ssec:quality}, for spectra with
$Q$=A,B,C,D,E we use $W_Q$=1,1,1,0.5,0 in computing redshifts,
$W_Q$=1,1,0.5,0,0 in computing dispersions, and $W_Q$=1,1,0.5,0,0 in
computing linestrengths.

(iii) {\em Run weighting:} we also apply a run-weighting $W_R$=0 to
exclude run 115, for reasons explained in \S\ref{ssec:setups}; all other
runs are given $W_R$=1.

The combined estimate $X$ is thus computed from the individual
measurements $X_i$ as the weighted mean
\begin{equation}
X = {\textstyle \sum_i} W_i X_i / {\textstyle \sum_i} W_i ~,
\label{eqn:combval}
\end{equation}
where $W_i = \Delta_i^{-2} W_{Qi} W_{Ri}$. The uncertainty in this
weighted mean is computed as
\begin{equation}
\Delta = ({\textstyle \sum_i} W_i)^{-1/2} ~.
\label{eqn:comberr}
\end{equation}
This procedure is used to obtain combined estimates of the redshift,
dispersion and linestrengths for each galaxy. We estimate the overall
quality $Q$ as the highest quality amongst the individual measurements
and obtain a combined estimate of the $S/N$ as
\begin{equation}
S/N = ({\textstyle \sum_i} (S/N)_i^{2} W_{Qi} W_{Ri})^{1/2} ~, 
\label{eqn:combsnr}
\end{equation}
using the same weightings as for the dispersions (except when the
overall quality is $Q$=D, when these weightings are omitted).

Table~\ref{tab:galtab} gives the combined estimates of the spectroscopic
parameters for each galaxy in the EFAR sample. The version of the table
presented here is abridged; the full table will be available upon
publication from NASA's Astrophysical Data Center (ADC) and from the
Centre de Donn\'{e}es astronomiques de Strasbourg (CDS). The table
lists: galaxy identification number (GIN), galaxy name, cluster
assignment number (CAN; see \S\ref{sec:clusass}), and the number of
spectra, redshifts, dispersions and \mgb\ and \mgtwo\ linestrengths
obtained for this object; then, for each of redshift, dispersion, \mgb\
and \mgtwo: the combined estimate, its estimated total error ($\Delta$)
and the weighted rms error from any repeat observations ($\delta$);
finally, the combined $S/N$ estimate and the overall quality parameter
(with an asterisk if the galaxy possesses emission lines). Note that
only objects with useful measurements are included; hence the lowest
quality class present in this table is $Q$=D, and the 7 galaxies with
only $Q$=E spectra (GINs 123, 284, 389, 448, 599, 637, 679) in
Table~\ref{tab:spectab} are omitted.

\begin{table*}
\centering
\caption{Spectroscopic parameters for the EFAR galaxies (abridged)}
\label{tab:galtab} 
\begin{tabular}{rlrcrrrrrrrrrrrrrl}
(1) & (2) & (3) & (4) & (5) & (6) & (7) & (8) & (9) & (10) & (11) & (12) & (13) & (14) & (15) & (16) & (17) & (18) \\
GIN & Galaxy & CAN & $N$ & $cz$ & $\Delta cz$ & $\delta cz$ & $\sigma$ & $\Delta\sigma$ & $\delta\sigma$ & 
\mgb & $\Delta$\mgb & $\delta$\mgb & \mgtwo & $\Delta$\mgtwo & $\delta$\mgtwo & $S/N$ & $Q$ \\
 & Name & & $s~z~\sigma~b~2$ & \multicolumn{3}{c}{\dotfill(\kms)\dotfill} & \multicolumn{3}{c}{\dotfill(\kms)\dotfill} &
\multicolumn{3}{c}{\dotfill(\AA)\dotfill} & \multicolumn{3}{c}{\dotfill(mag)\dotfill} & & \vspace*{6pt} \\
  1 & A76~A     &   1 & 1~1~1~1~1 & 11296 &  53 & --- & 262 &  53 & --- &  5.20 &  0.57 &   --- & 0.353 & 0.024 &   --- &  21.4 & A  \\
  2 & A76~B     &   1 & 1~1~1~1~1 & 11317 &  49 & --- & 224 &  49 & --- &  4.76 &  0.62 &   --- & 0.326 & 0.024 &   --- &  18.8 & B  \\
  3 & A76~C     &   1 & 1~1~1~1~1 & 11973 &  51 & --- & 217 &  50 & --- &  5.17 &  0.70 &   --- & 0.392 & 0.025 &   --- &  17.8 & A  \\
  4 & A76~D     &   1 & 1~1~1~1~1 & 12184 &  33 & --- & 316 & 150 & --- &  6.18 &  1.21 &   --- & 0.326 & 0.041 &   --- &  10.5 & C  \\
  5 & A76~E     &   1 & 2~2~2~2~2 & 12189 &  14 &  47 & 184 &  10 &  23 &  6.26 &  0.18 &  0.09 & 0.295 & 0.010 & 0.018 &  46.7 & B  \\
  6 & A76~F     &   1 & 1~1~1~1~1 & 12371 &  30 & --- & 307 &  85 & --- &  5.29 &  0.86 &   --- & 0.331 & 0.028 &   --- &  17.6 & B  \\
  8 & A85~A     &   2 & 2~2~1~1~1 & 16604 &  35 &  49 & 290 &  53 & --- &  5.14 &  0.61 &   --- & 0.344 & 0.022 &   --- &  18.7 & B  \\
  9 & A85~B     &   2 & 1~1~1~1~0 & 17349 &  35 & --- & 436 &  47 & --- &  5.74 &  0.25 &   --- &   --- &   --- &   --- &  37.6 & A  \\
 10 & A85~C     & 102 & 1~1~0~0~0 & 22837 &  71 & --- & --- & --- & --- &   --- &   --- &   --- &   --- &   --- &   --- &  33.9 & D  \\
 11 & A85~1     &   2 & 1~1~1~1~1 & 15112 &  22 & --- & 195 &  19 & --- &  3.82 &  0.42 &   --- & 0.253 & 0.018 &   --- &  25.8 & B  \\
 12 & A85~2     &   2 & 1~1~1~1~1 & 16264 &  31 & --- & 294 &  37 & --- &  5.86 &  0.37 &   --- & 0.337 & 0.019 &   --- &  28.3 & B  \\
 13 & A119~A    &   3 & 3~3~3~3~2 & 11459 &  15 &  17 & 297 &  15 &  13 &  5.04 &  0.16 &  0.31 & 0.316 & 0.009 & 0.010 &  56.4 & A  \\
 14 & A119~B    &   3 & 2~2~2~2~2 & 13330 &  20 &  42 & 278 &  21 &  10 &  4.90 &  0.22 &  0.21 & 0.358 & 0.010 & 0.011 &  36.5 & A  \\
 15 & A119~C    &   3 & 2~2~2~2~1 & 13500 &  16 &  11 & 253 &  17 &   6 &  5.57 &  0.20 &  0.14 & 0.295 & 0.011 &   --- &  38.5 & A  \\
 16 & A119~D    &   3 & 3~3~1~1~1 & 14982 &  16 &   9 & 104 &  18 & --- &  3.30 &  0.52 &   --- & 0.151 & 0.023 &   --- &  14.7 & C* \\
 17 & A119~E    &   3 & 2~2~2~2~1 & 12798 &  14 &   9 & 248 &  14 &   4 &  5.37 &  0.16 &  0.35 & 0.326 & 0.010 &   --- &  45.1 & A  \\
 18 & A119~F    &   3 & 2~2~2~2~1 & 13018 &  12 &  14 & 175 &  13 &  33 &  4.88 &  0.28 &  0.11 & 0.245 & 0.013 &   --- &  32.1 & B  \\
 19 & A119~G    &   3 & 2~1~1~1~0 & 13457 &  20 & --- & 267 &  21 & --- &  4.78 &  0.22 &   --- &   --- &   --- &   --- &  36.7 & A  \\
 21 & A119~I    &   3 & 1~1~1~1~1 & 13271 &  22 & --- & 225 &  20 & --- &  4.85 &  0.41 &   --- & 0.326 & 0.016 &   --- &  31.0 & A  \\
 22 & A119~J    &   3 & 1~1~1~1~0 & 13520 &  22 & --- & 219 &  25 & --- &  4.55 &  0.35 &   --- &   --- &   --- &   --- &  22.3 & B  \\
 23 & A119~1    & 103 & 1~1~1~1~0 &  4127 &  19 & --- &  92 &  54 & --- &  0.50 &  1.12 &   --- &   --- &   --- &   --- &   8.1 & C* \\
 24 & A119~2    &   3 & 2~1~1~1~0 & 12346 &  17 & --- & 100 &  20 & --- &  5.13 &  0.39 &   --- &   --- &   --- &   --- &  20.4 & B  \\
 25 & J3~A      &   4 & 1~1~1~1~1 & 14453 &  34 & --- & 333 & 114 & --- &  6.25 &  0.77 &   --- & 0.329 & 0.026 &   --- &  15.2 & B  \\
 26 & J3~B      &   4 & 1~1~1~1~0 & 13519 &  19 & --- & 231 &  16 & --- &  4.93 &  0.26 &   --- &   --- &   --- &   --- &  39.5 & A  \\
 27 & J3~C      &   4 & 1~1~1~1~1 & 13770 &  23 & --- & 163 &  57 & --- &  4.33 &  1.26 &   --- & 0.224 & 0.040 &   --- &  11.2 & C  \\
 28 & J3~D      &   4 & 2~2~2~2~0 & 14610 &  15 & 258 & 231 &  13 &  37 &  5.18 &  0.20 &  0.29 &   --- &   --- &   --- &  47.6 & A  \\
 31 & J4~A      &   5 & 1~1~1~1~1 & 12074 &  26 & --- & 185 &  75 & --- &  5.35 &  1.11 &   --- & 0.387 & 0.043 &   --- &   9.3 & C  \\
 32 & J4~B      &   5 & 1~1~1~1~1 & 12090 &  33 & --- & 302 & 148 & --- &  4.11 &  1.56 &   --- & 0.261 & 0.047 &   --- &   9.4 & C  \\
 33 & J4~C      & 105 & 1~1~1~1~1 & 17154 &  36 & --- & 358 &  48 & --- &  6.20 &  0.32 &   --- & 0.355 & 0.019 &   --- &  30.0 & A  \\
 36 & A147~A    &   6 & 4~4~4~4~4 & 12771 &  11 &  19 & 244 &   9 &  12 &  4.88 &  0.13 &  0.14 & 0.289 & 0.007 & 0.008 &  89.9 & A  \\
 37 & A147~B    &   6 & 4~4~4~4~4 & 13119 &  11 &  15 & 316 &  10 &  20 &  4.68 &  0.11 &  0.30 & 0.304 & 0.006 & 0.017 & 104.2 & A  \\
 38 & A147~C    &   6 & 3~3~3~3~3 & 13156 &  13 &   8 & 247 &  12 &   7 &  5.22 &  0.15 &  0.22 & 0.305 & 0.008 & 0.007 &  66.8 & A  \\
 39 & A147~D    &   6 & 3~3~3~3~3 & 13444 &  12 &   2 & 185 &   9 &  15 &  4.98 &  0.18 &  0.28 & 0.294 & 0.008 & 0.004 &  67.8 & A  \\
 40 & A147~E    &   6 & 3~3~3~3~3 & 13049 &  10 &  10 & 176 &   8 &   7 &  4.58 &  0.14 &  0.12 & 0.267 & 0.007 & 0.002 &  76.1 & A  \\
 41 & A147~F    &   6 & 2~2~2~2~2 & 11922 &  12 &   5 &  87 &  10 &   4 &  3.65 &  0.27 &  0.02 & 0.195 & 0.012 & 0.002 &  37.9 & B  \\
 42 & A147~1    &   6 & 3~3~3~3~3 & 12832 &  11 &  29 & 148 &   9 &   3 &  4.35 &  0.16 &  0.09 & 0.252 & 0.008 & 0.013 &  65.8 & A  \\
 43 & A160~A    &   7 & 2~2~2~2~1 & 11401 &  15 &   3 & 181 &  17 &  19 &  3.86 &  0.26 &  0.04 & 0.250 & 0.017 &   --- &  35.1 & A  \\
 44 & A160~B    & 107 & 3~3~1~1~1 & 18258 &  24 &  17 & 192 &  21 & --- &  6.61 &  0.30 &   --- & 0.289 & 0.022 &   --- &  26.3 & B  \\
 45 & A160~C    &   7 & 1~1~1~1~0 & 12380 &  34 & --- & 412 &  58 & --- &  4.61 &  0.37 &   --- &   --- &   --- &   --- &  27.8 & A  \\
 46 & A160~D    & 107 & 3~3~0~0~0 & 18271 &  41 &  84 & --- & --- & --- &   --- &   --- &   --- &   --- &   --- &   --- &  36.1 & D  \\
 47 & A160~E    &   7 & 4~4~4~4~2 & 14056 &  13 &  24 & 226 &  16 &  12 &  5.01 &  0.23 &  0.26 & 0.293 & 0.017 & 0.014 &  39.2 & A  \\
 48 & A160~F    &   7 & 3~3~3~3~2 & 13657 &  14 &  16 & 176 &  18 &  27 &  5.18 &  0.24 &  0.43 & 0.307 & 0.013 & 0.009 &  33.9 & A  \\
 49 & A160~G    &   7 & 4~4~4~4~2 & 13137 &  13 &  25 & 196 &  20 &  22 &  4.82 &  0.25 &  0.39 & 0.293 & 0.015 & 0.056 &  32.4 & B  \\
 50 & A160~H    &   7 & 1~1~1~1~0 & 13589 &  21 & --- & 195 &  23 & --- &  5.18 &  0.30 &   --- &   --- &   --- &   --- &  21.6 & A  \\
 51 & A160~I    & 107 & 3~3~0~0~0 & 18643 &  41 &  30 & --- & --- & --- &   --- &   --- &   --- &   --- &   --- &   --- &  43.9 & D  \\
 52 & A160~J    &   7 & 4~4~4~4~1 & 11254 &   9 &  13 & 145 &  10 &  15 &  2.74 &  0.20 &  0.32 & 0.217 & 0.041 &   --- &  46.1 & A* \\
 53 & A160~1    & 107 & 2~2~0~0~0 & 18108 &  50 & 103 & --- & --- & --- &   --- &   --- &   --- &   --- &   --- &   --- &  27.3 & D* \\
 54 & A160~2    & 107 & 1~1~0~0~0 & 18201 &  71 & --- & --- & --- & --- &   --- &   --- &   --- &   --- &   --- &   --- &  21.7 & D  \\
 56 & A168~A    & 108 & 1~1~1~1~0 &  5299 &  19 & --- & 265 &  16 & --- &  4.35 &  0.26 &   --- &   --- &   --- &   --- &  51.7 & A  \\
 57 & A168~B    & 108 & 1~1~1~1~0 &  5253 &  23 & --- & 310 &  25 & --- &  5.29 &  0.30 &   --- &   --- &   --- &   --- &  43.6 & A  \\
\end{tabular}
\vspace*{6pt} \\ \parbox{\textwidth}{This is an abridged version of this
table; the full table will be available upon publication from NASA's
Astrophysical Data Center (ADC) and from the Centre de Donn\'{e}es
astronomiques de Strasbourg (CDS). The columns of this table give:
(1)~galaxy identification number (GIN); (2)~galaxy name; (3)~the cluster
assignment number (CAN); (4)~the number of spectra $N_s$, redshifts
$N_z$, dispersions $N_{\sigma}$, \mgb\ linestrengths $N_b$ and \mgtwo\
linestrengths $N_2$ obtained for this object; then the combined
estimate, its estimated total error ($\Delta$) and the weighted rms
error from any repeat observations ($\delta$) for each of
(5--7)~redshift, (8--10)~dispersion, (11--13)~\mgb\ linestrength and
(14--16)~\mgtwo\ linestrength; (17)~the combined $S/N$ estimate; and
(18)~the overall quality parameter (with an asterisk if the galaxy
possesses emission lines). Only objects with useful measurements are
included; hence the lowest quality class present in this table is $Q$=D,
and the 7 galaxies with only $Q$=E spectra (GINs 123, 284, 389, 448,
599, 637, 679) are omitted.}
\end{table*}

\begin{figure}
\plotone{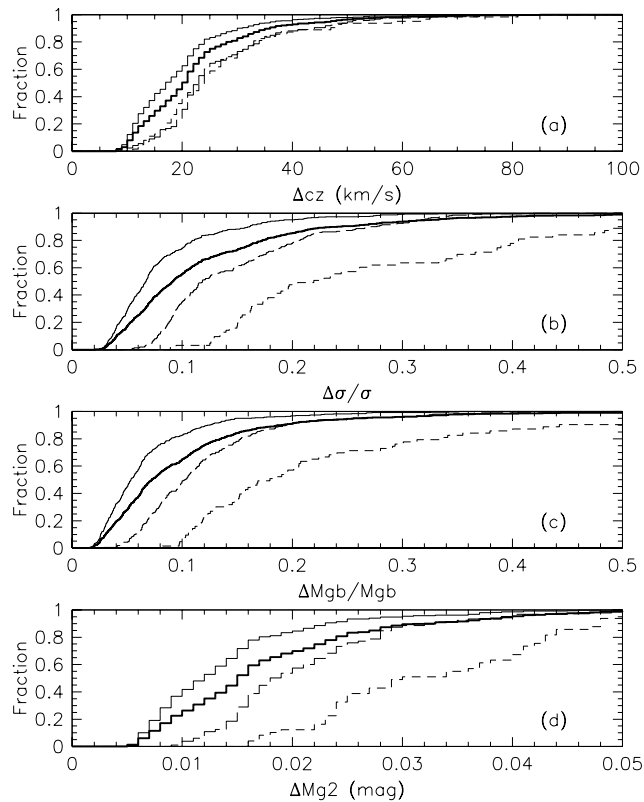}
\caption{The cumulative distributions of the total estimated errors in
the combined measurements of redshift, velocity dispersion, \mgb\ and
\mgtwo\ for each galaxy. The distributions for quality classes A, B and
C are shown as full, long-dashed and short-dashed lines respectively;
the overall distribution is shown as the thick full line. (a)~The
distribution of combined errors in $cz$; (b)~combined relative errors in
$\sigma$; (c)~combined relative errors in \mgb; (d)~combined errors in
\mgtwo.
\label{fig:errdist2}}
\end{figure}

The cumulative distributions of uncertainties in the combined results
are shown in Figure~\ref{fig:errdist2}, both for the entire dataset and
for quality classes A, B and C separately. The error distributions can
be quantitatively characterised by their 50\% and 90\% points, which are
listed in Table~\ref{tab:typerrs2}. The overall median error in redshift
is 20\kms, the median relative errors in dispersion and \mgb\ are 9.1\%
and 7.2\%, and the median error in \mgtwo\ is 0.015~mag. For the whole
sample, and for quality classes A and B, the median errors in the
combined measurements are smaller than the median errors in the
individual measurements, as one expects. However for dispersion, \mgb\
and \mgtwo\ the errors are larger for quality class C and at the 90th
percentile; this results from assigning a quality weighting of 0.5
to $Q$=C when combining the individual measurements of these quantities.

The uncertainties listed in Table~\ref{tab:galtab} represent the best
estimates of the total errors in the parameters for each galaxy. However
it must be emphasised that they are {\em not} independent of each other,
as the run correction errors are correlated across all measurements from
a run. To properly simulate the joint distribution of some parameter for
the whole dataset, one must first generate realisations of the run
correction errors (drawn from Gaussians with standard deviations given
by the uncertainties listed in Table~\ref{tab:runcorr}) and the
individual measurement errors (drawn from Gaussians with standard
deviations given by the uncertainties listed Table~\ref{tab:spectab}).
For each individual measurement, one must add the realisation of its
measurement error and the realisation of the appropriate run correction
error (the same for all measurements in a given run) to the measured
value of the parameter. The resulting realisations of the individual
measurements are finally combined using the recipe described above to
yield a realisation of the value of the parameter for each galaxy in the
dataset.

\begin{table}
\centering
\caption{The distribution of errors per galaxy}
\label{tab:typerrs2}
\begin{tabular}{ccccccccc}
$Q$ & \multicolumn{2}{c}{$\Delta cz$ (km/s)} & 
\multicolumn{2}{c}{$\Delta\sigma/\sigma$} & 
\multicolumn{2}{c}{$\Delta$\mgb/\mgb} &
\multicolumn{2}{c}{$\Delta$\mgtwo\ (mag)} \\
    & 50\% & 90\% &  50\% &  90\% &  50\% &  90\% &  50\% &  90\% \\
All & 20   & 36   & 0.091 & 0.240 & 0.072 & 0.188 & 0.015 & 0.032 \\
A   & 17   & 30   & 0.067 & 0.161 & 0.053 & 0.120 & 0.012 & 0.023 \\
B   & 23   & 44   & 0.118 & 0.270 & 0.103 & 0.194 & 0.018 & 0.033 \\
C   & 24   & 47   & 0.220 & 0.507 & 0.180 & 0.425 & 0.030 & 0.048 \\
\end{tabular}
\end{table}

\begin{figure}
\plotone{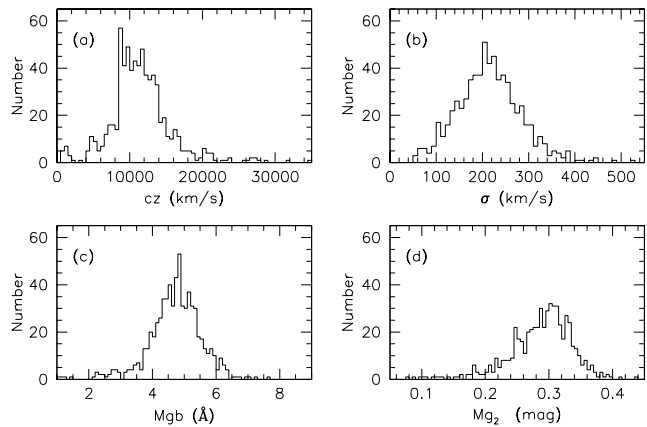}
\caption{The distributions of (a)~redshift, (b)~velocity dispersion,
(c)~\mgb\ linestrength, and (d)~\mgtwo\ linestrength for the galaxies in
the EFAR sample.
\label{fig:specpar}}
\end{figure}

\begin{figure}
\plotone{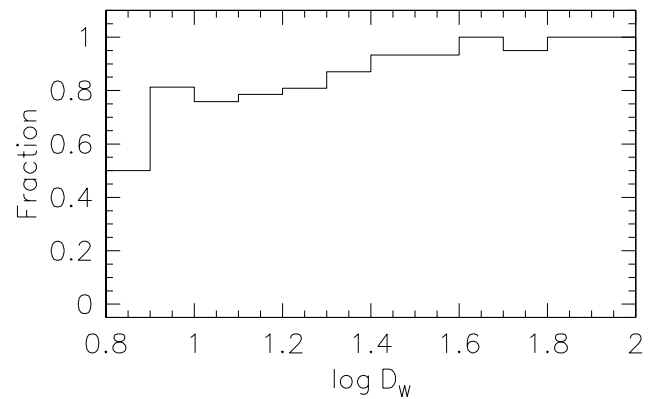}
\caption{The fraction of programme objects for which we measured a
velocity dispersion as a function of the logarithm of the selection
diameter $D_W$ (in arcsec).
\label{fig:dwsig}}
\end{figure}

The distributions of redshift, velocity dispersion, \mgb\ and \mgtwo\
for the galaxies in the EFAR sample are displayed in
Figure~\ref{fig:specpar}. The galaxies for which we measured velocity
dispersions are only a subset of our sample of programme galaxies
(629/743), and represent a refinement of the sample selection criteria.
Figure~\ref{fig:dwsig} shows the fraction of programme galaxies with
measured dispersions as a function of the galaxy diameter $D_W$ on which
the selection function of the programme galaxy sample is defined. There
is a steady decline in the fraction of the sample for which usable
dispersions were measured, from 100\% for the largest galaxies (with
$D_W \gs 40$~arcsec) to about 75\% for the smallest (with 8~arcsec~$\ls
D_W \ls$~15~arcsec; there are only 3 programme galaxies with $D_W
<$~8~arcsec). This additional selection effect must be allowed for when
determining Fundamental Plane distances.

\subsection{Internal and External Comparisons}
\label{ssec:compare}

\begin{figure}
\plotone{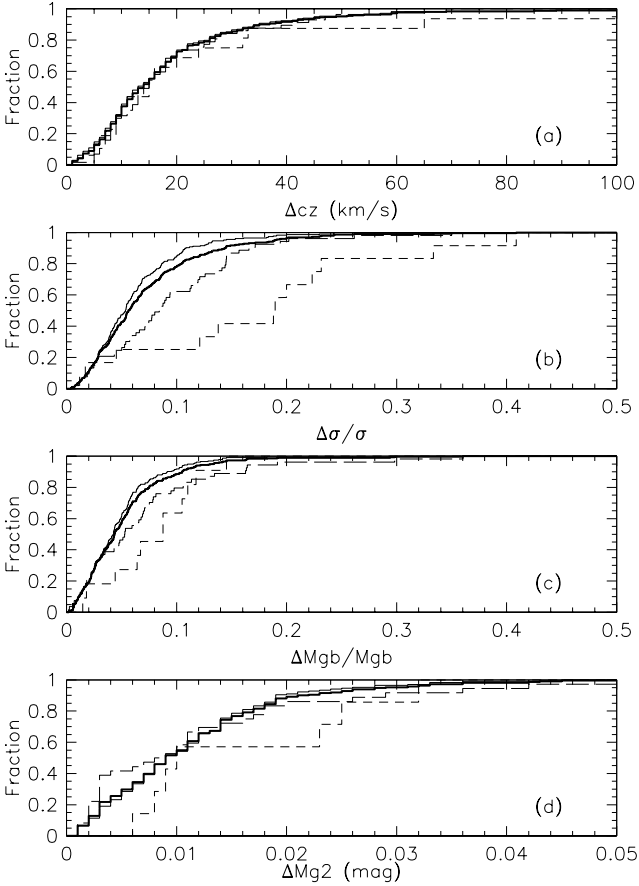}
\caption{The cumulative distributions of the rms errors from repeat
measurements of redshift, velocity dispersion, \mgb\ and \mgtwo. The
distributions for spectral types A, B and C are shown as full,
long-dashed and short-dashed lines respectively; the overall
distribution is shown as the thick full line. (a)~The distribution of
rms errors in $cz$ in \kms; (b)~relative rms errors in $\sigma$;
(c)~relative rms errors in \mgb; (d)~rms errors in \mgtwo.
\label{fig:errdist3}}
\end{figure}

One of the strengths of our spectroscopic sample is the high fraction
of objects with repeat observations: there are 375 galaxies with a
single dispersion measurement, 160 with two measurements and 141 with
three or more measurements. Figure~\ref{fig:errdist3} shows the
cumulative distributions of rms errors in redshift, dispersion, \mgb\
and \mgtwo\ obtained from these repeat observations. The detailed
internal comparisons made possible by these repeat measurements have
been used to establish the run corrections (\S\ref{ssec:combruns}) and
to calibrate the estimated errors (\S\ref{ssec:caliberr}). The latter
process ensured that the estimated errors were statistically
consistent with the rms errors of the repeat measurements.

\begin{figure}
\plotone{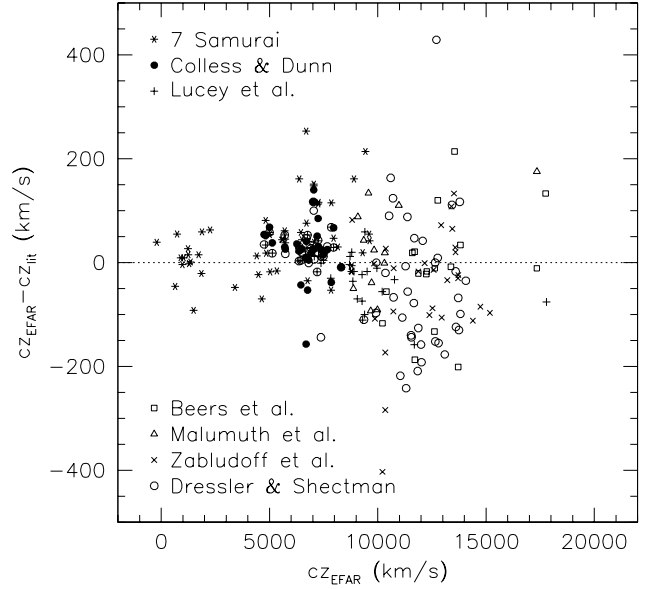}
\caption{Differences between EFAR redshifts and those from various
sources in the literature.
\label{fig:compcz}}
\end{figure}

We also make external comparisons of our measurements with the work of
other authors. The EFAR redshifts are compared in
Figure~\ref{fig:compcz} with redshifts given in the literature by the
7~Samurai (Davies \etal\ 1987), Dressler \& Shectman (1988), Beers
\etal\ (1991), Malumuth \etal\ (1992), Zabludoff \etal\ (1993), Colless
\& Dunn (1996) and Lucey \etal\ (1997). Only 11 of the 256 comparisons
give redshift differences greater than 300\kms: in 6 cases the EFAR
redshift is confirmed either by repeat measurements or other published
measurements; in the remaining 5 cases the identification of the galaxy
in question is uncertain in the literature. For the 245 cases where the
redshift difference is less than 300\kms, there is no significant
velocity zeropoint error and the rms scatter is 85\kms. Since our repeat
measurements show much smaller errors (90\% are less than 36\kms), most
of this scatter must arise in the literature data, some of which were
taken at lower resolution or $S/N$ than our data.

\begin{figure}
\plotone{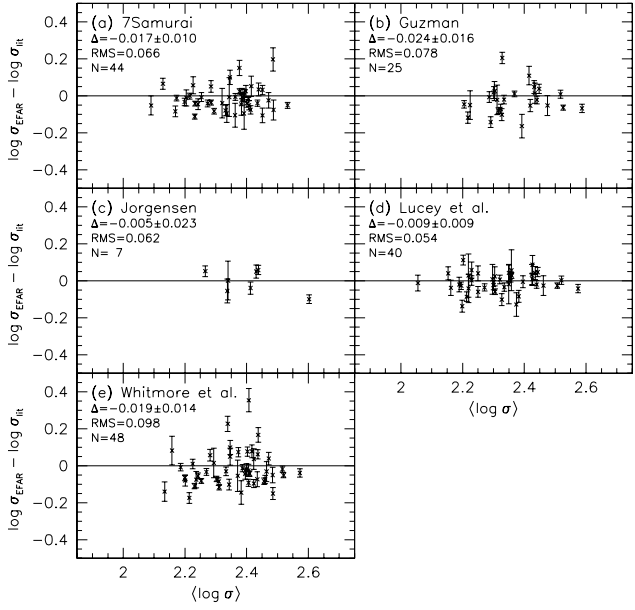}
\caption{Comparisons of EFAR dispersions with those from various sources
in the literature: (a)~Davies \etal\ (1987), (b)~Guzm\'{a}n (1993),
(c)~J{\o}rgensen (1997), (d) Lucey \etal\ (1997), and (e)~Whitmore
\etal\ (1985). In each case the mean difference,
$\Delta=\langle\log\sigma_{EFAR}-\log\sigma_{lit}\rangle$, and its
standard error are indicated, along with the rms scatter and the number
of galaxies in the comparison.
\label{fig:compsig}}
\end{figure}

Figure~\ref{fig:compsig} compares the EFAR dispersions with published
dispersions from the work of the 7~Samurai (Davies \etal\ 1987),
Guzm\'{a}n (1993), J{\o}rgensen \etal\ (1995) and Lucey \etal\ (1997),
and the compilation of earlier measurements by Whitmore \etal\ (1985).
Note that we do not compare to the more recent compilation by McElroy
(1995), since its overlap with our sample is essentially just the sum of
above sources. The mean differences,
$\Delta=\log\sigma_{EFAR}-\log\sigma_{lit}$, and their standard errors
are indicated on the figure; none of these scale differences is larger
than 6\% and in fact all five comparisons are consistent with zero scale
error at the 2$\sigma$ level or better. The rms scatter in these
comparisons is significantly greater than the errors in our dispersion
measurements, implying that in general the literature measurements have
larger errors and/or that there are unaccounted-for uncertainties in the
comparison.

\begin{figure}
\plotone{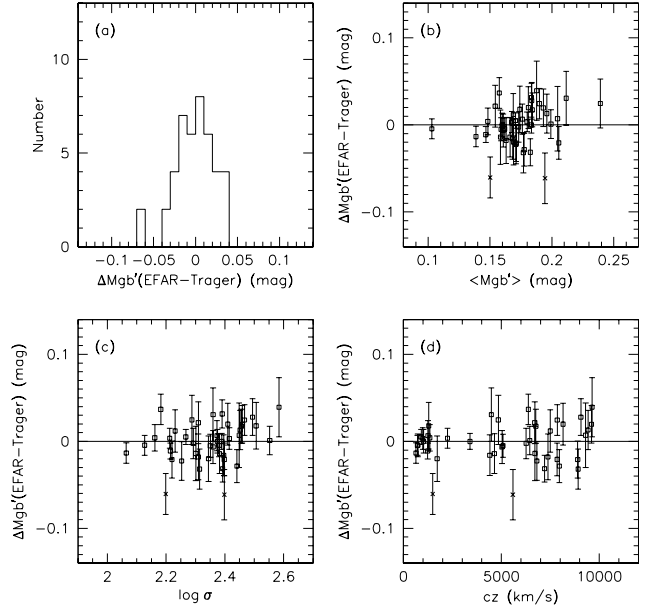}
\caption{The residual differences in \mgbp\ linestrengths from EFAR and
Trager \etal\ (1998) {\em after} applying the zeropoint corrections
discussed in the text: (a)~the distribution of residuals; (b)~the
residuals as a function of \mgbp; (c)~the residuals as a function of
$\log\sigma$; (d)~the residuals as a function of redshift. Outliers
excluded from the determination of the zeropoint corrections are shown
as crosses.
\label{fig:mgbcomp}}
\vspace*{0.5cm}
\end{figure}

\begin{figure}
\plotone{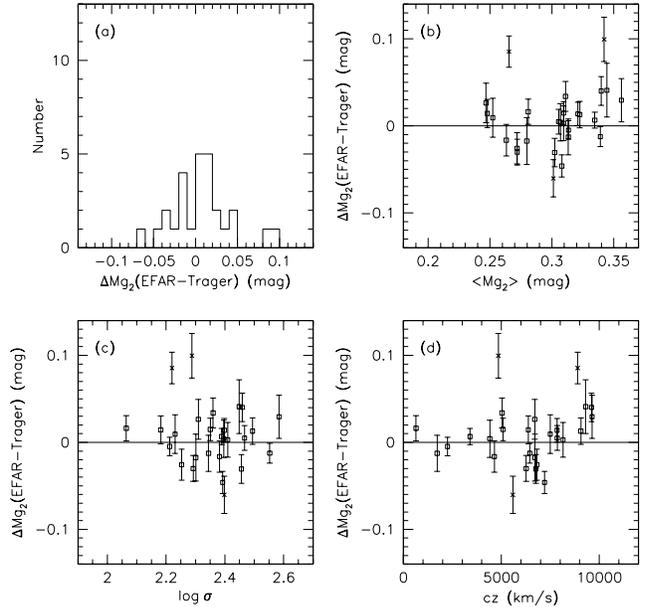}
\caption{The residual differences in \mgtwo\ linestrengths from EFAR and
Trager \etal\ (1998) {\em after} applying the zeropoint corrections
discussed in the text: (a)~the distribution of residuals; (b)~the
residuals as a function of \mgtwo; (c)~the residuals as a function of
$\log\sigma$; (d)~the residuals as a function of redshift. Outliers
excluded from the determination of the zeropoint corrections are shown
as crosses.
\label{fig:mg2comp}}
\vspace*{0.5cm}
\end{figure}

We determine the zeropoint calibration of our linestrength measurements
with respect to the Lick system (see \S\ref{ssec:indices}) by comparing
our \mgbp\ and \mgtwo\ linestrengths to measurements for the same
galaxies given by Trager \etal\ (1998). We find that slightly different
calibrations are needed for objects with different redshifts, the result
of slight variations in the non-linear continuum shape as the spectra
are redshifted with respect to the instrument response and the sky
background (see \S\ref{ssec:indices}). Good agreement with Trager \etal\
is obtained if we use different zeropoints for galaxies with redshifts
above and below $cz$=3000\kms\ (although there are no objects in the
comparison at $cz$$>$10000\kms). Excluding a few outliers, we find
weighted mean differences between the EFAR and Trager \etal\
linestrengths of $\langle\Delta\mgbp\rangle$=$-0.022$~mag and
$\langle\Delta\mgtwo\rangle$=$-0.083$~mag for $cz$$<$3000\kms, and
$\langle\Delta\mgbp\rangle$=$-0.008$~mag and
$\langle\Delta\mgtwo\rangle$=$-0.028$~mag for $cz$$>$3000\kms.
Subtracting these zeropoint corrections gives the final,
fully-corrected, linestrength measurements as listed in
Tables~\ref{tab:spectab} \&~\ref{tab:galtab}. Figures~\ref{fig:mgbcomp}
\&~\ref{fig:mg2comp} show the residual differences between the EFAR and
Trager \etal\ linestrength measurements after applying these zeropoint
corrections. The rms scatter is 0.019~mag in \mgbp\ for the 41 objects
in common, and 0.023~mag in \mgtwo\ for the 24 objects in common. There
is no statistically significant trend with linestrength, velocity
dispersion or redshift remaining in the residuals after these zeropoint
corrections are applied.

\begin{figure}
\plotone{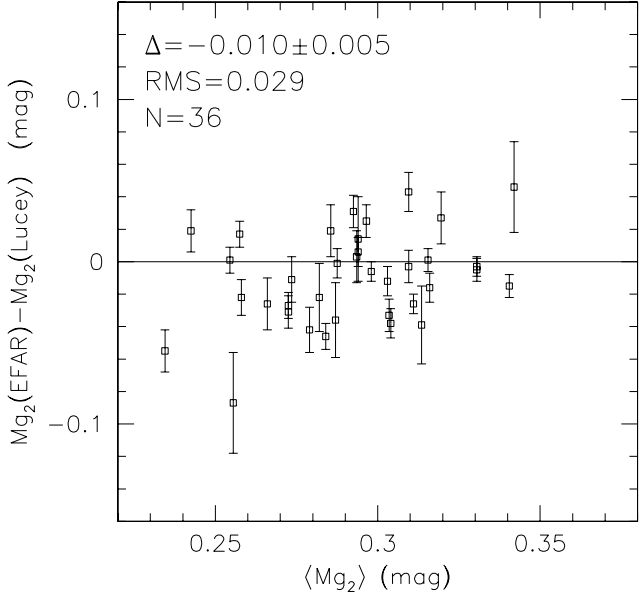}
\caption{Comparisons of \mgtwo\ linestrengths obtained by EFAR and
Lucey \etal\ (1997). The mean difference, $\Delta$ = \mgtwo(EFAR) $-$
\mgtwo(Lucey), and its standard error are indicated, along with the
rms scatter and the number of galaxies in the comparison.
\label{fig:compmg2}}
\vspace*{0.5cm}
\end{figure}

Figure~\ref{fig:compmg2} compares our calibrated \mgtwo\ linestrengths
to those obtained in A2199, A2634 and Coma by Lucey \etal\ (1997). The
overall agreement for the 36 objects in common is very good, with a
statistically non-significant zeropoint offset and an rms scatter of
0.029~mag, similar to that found in the comparison with Trager \etal

\begin{figure}
\vspace*{0.9cm}
\plotone{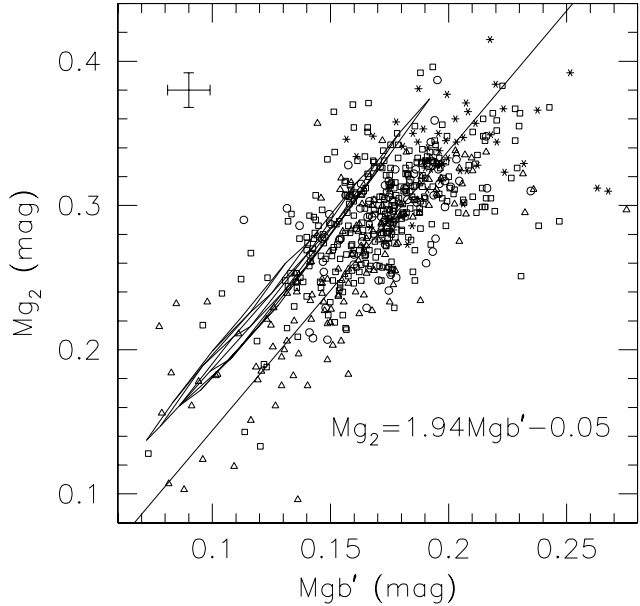}
\caption{The relation between \mgbp\ and \mgtwo\ and its maximum
likelihood fit. Ellipticals are marked by circles, E/S0s by squares, cDs
by asterisks and spirals by triangles. Typical estimated errors are
shown in the top left corner. The relation between \mgbp\ and \mgtwo\ as
a function of age and metallicity, as predicted by Worthey (1994), is
shown as the grid lying parallel to, but offset from, the data.
\label{fig:mgbmg2}}
\vspace*{0.5cm}
\end{figure}

The relation between the measured \mgbp\ and \mgtwo\ linestrengths for
all the galaxies in the EFAR sample is shown in Figure~\ref{fig:mgbmg2}.
We fit this relation using a maximum likelihood technique which accounts
for both measurement errors and selection effects (Saglia \etal\ 1998,
in preparation; Paper~VI). We find
\begin{equation}
\mgtwo = 1.94 \mgbp - 0.05 ~,
\label{eqn:mgbmg2}
\end{equation}
with a perpendicular rms residual of 0.019~mag (corresponding to an rms
of 0.041~mag in \mgtwo, or 0.021~mag in \mgbp). The relation is the same
if we fit ellipticals, E/S0s, cDs or spirals separately. This relation
is similar to those derived by Burstein \etal\ (1984) and J{\o}rgensen
(1997). We can therefore use \mgbp\ as a predictor of \mgtwo\ (albeit
with larger uncertainties) for those cases where \mgtwo\ cannot be
measured directly.

Also shown in Figure~\ref{fig:mgbmg2} is the predicted relation between
\mgbp\ and \mgtwo\ as a function of age and metallicity given by Worthey
(1994). His models correctly predict the slope of the relation, but are
offset by $-0.025$~mag in \mgbp (or by $+0.05$~mag in \mgtwo),
indicating a difference in the model's zeropoint calibration for one or
both indices.

\section{CLUSTER ASSIGNMENTS}
\label{sec:clusass}

The correct assignment of galaxies to clusters (or groups) is crucial to
obtaining reliable redshifts and distances for the EFAR cluster
sample. We also need to increase the precision of the cluster redshifts
in order to minimise uncertainties in the clusters' peculiar velocities.
To achieve these goals we merged the EFAR redshifts with redshifts for
all galaxies in ZCAT (Huchra \etal, 1992; version of 1997 May 29) which
lie within 3\Mpc\ (2 Abell radii) of each nominal EFAR cluster centre
(see Table~1 of Paper~I). We then examined the redshift distributions of
the combined sample in order to distinguish groups, clusters and field
galaxies along the line of sight to a nominal EFAR `cluster'. We also
considered the distribution of galaxies on the sky before assigning the
EFAR galaxies to specific groupings.

\begin{figure*}
\plotfull{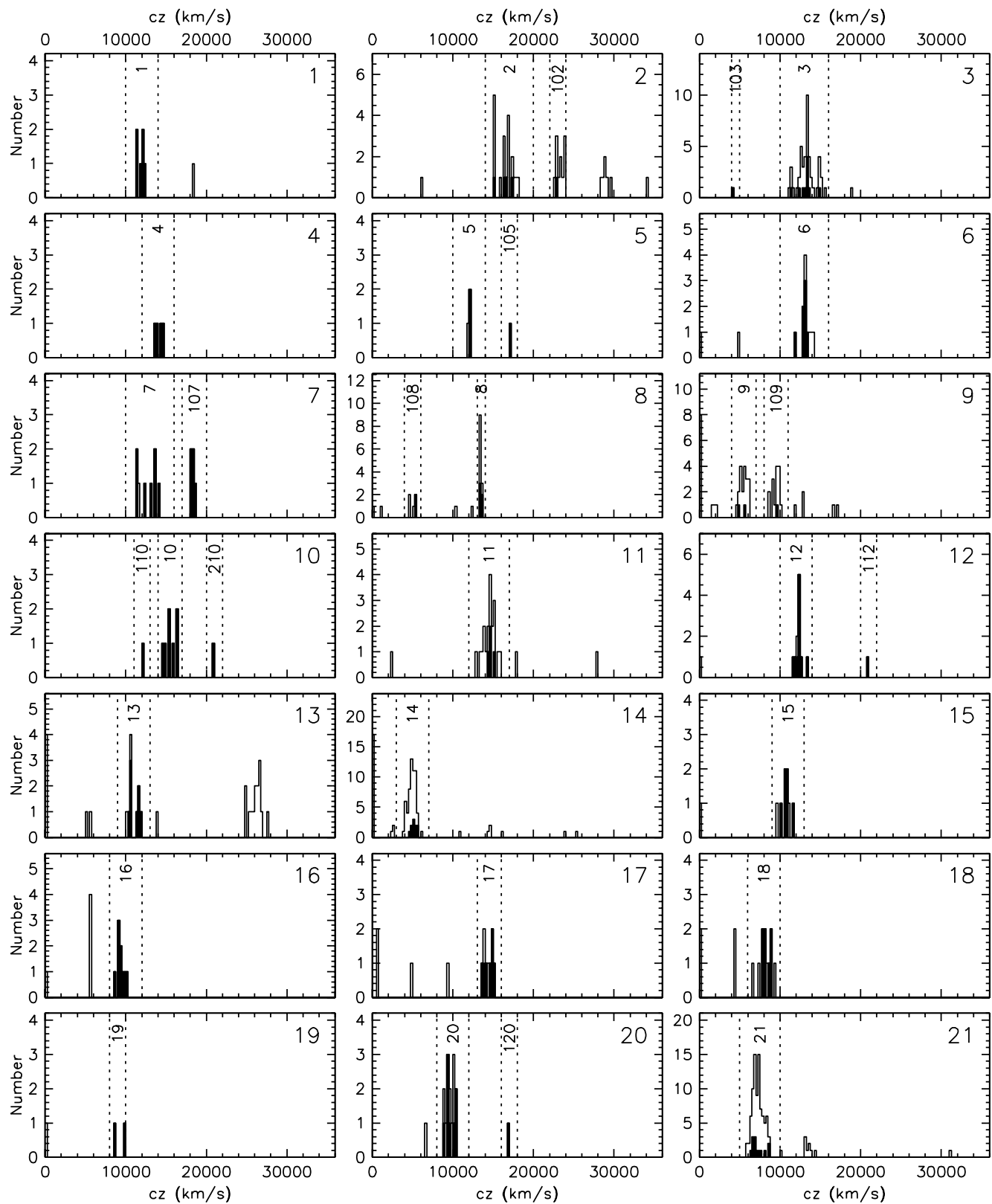}
\caption{The redshift distributions of galaxies within 3\Mpc\ of each
nominal EFAR cluster using the EFAR and ZCAT data. Each distribution is
labelled at top right by the nominal cluster ID number (CID). The solid
histogram shows the distribution of EFAR galaxies; the open histogram
shows the extra ZCAT galaxies. The groupings adopted have boundaries in
redshift marked by dotted lines and are labelled by their cluster
assignment number (CAN). Clusters without numbers and boundaries contain
no EFAR galaxies.
\label{fig:cluz}}
\end{figure*}
\addtocounter{figure}{-1}
\begin{figure*}
\plotfull{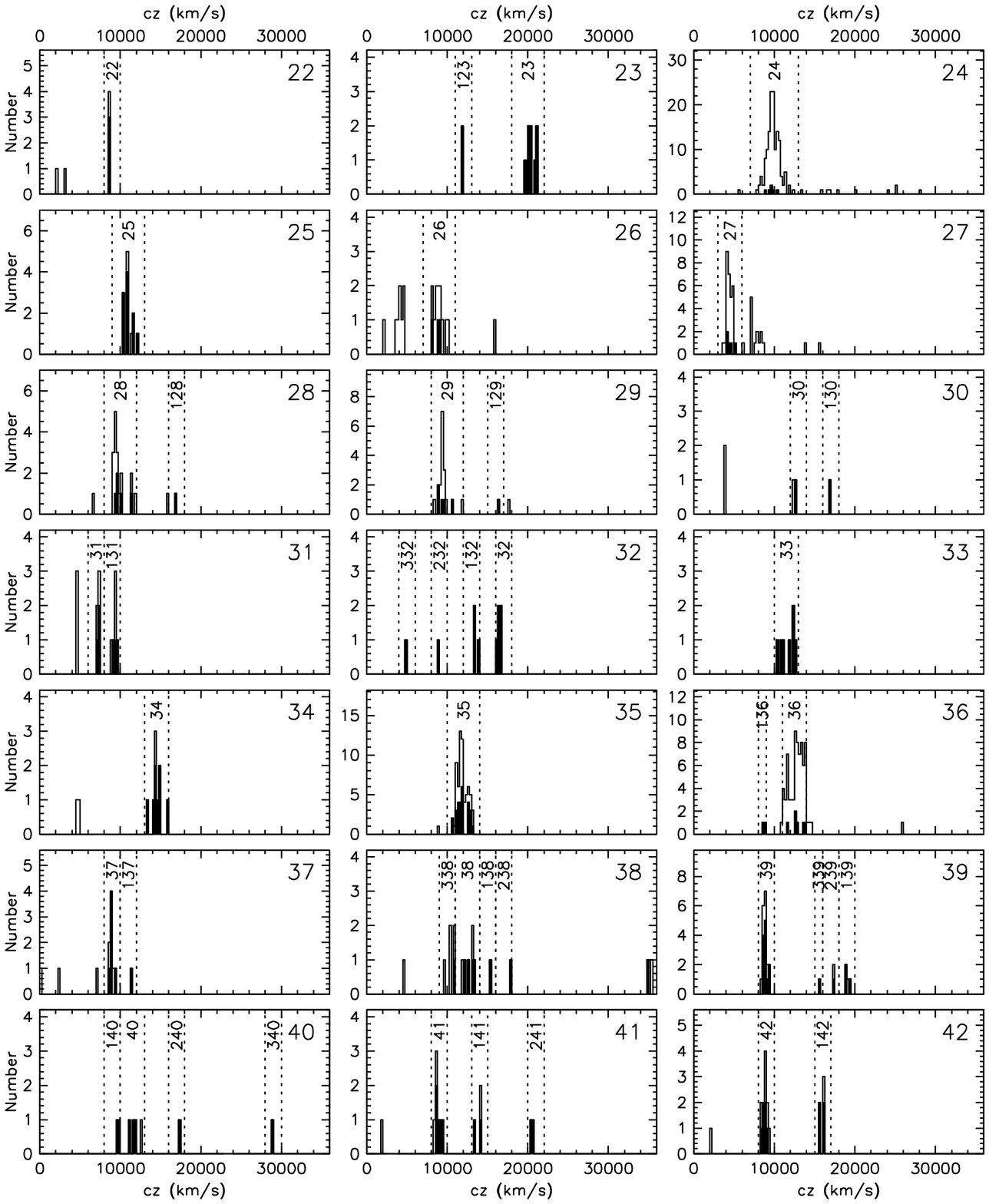}
\caption{{\em (ctd)}}
\vspace*{1cm}
\end{figure*}
\addtocounter{figure}{-1}
\begin{figure*}
\plotfull{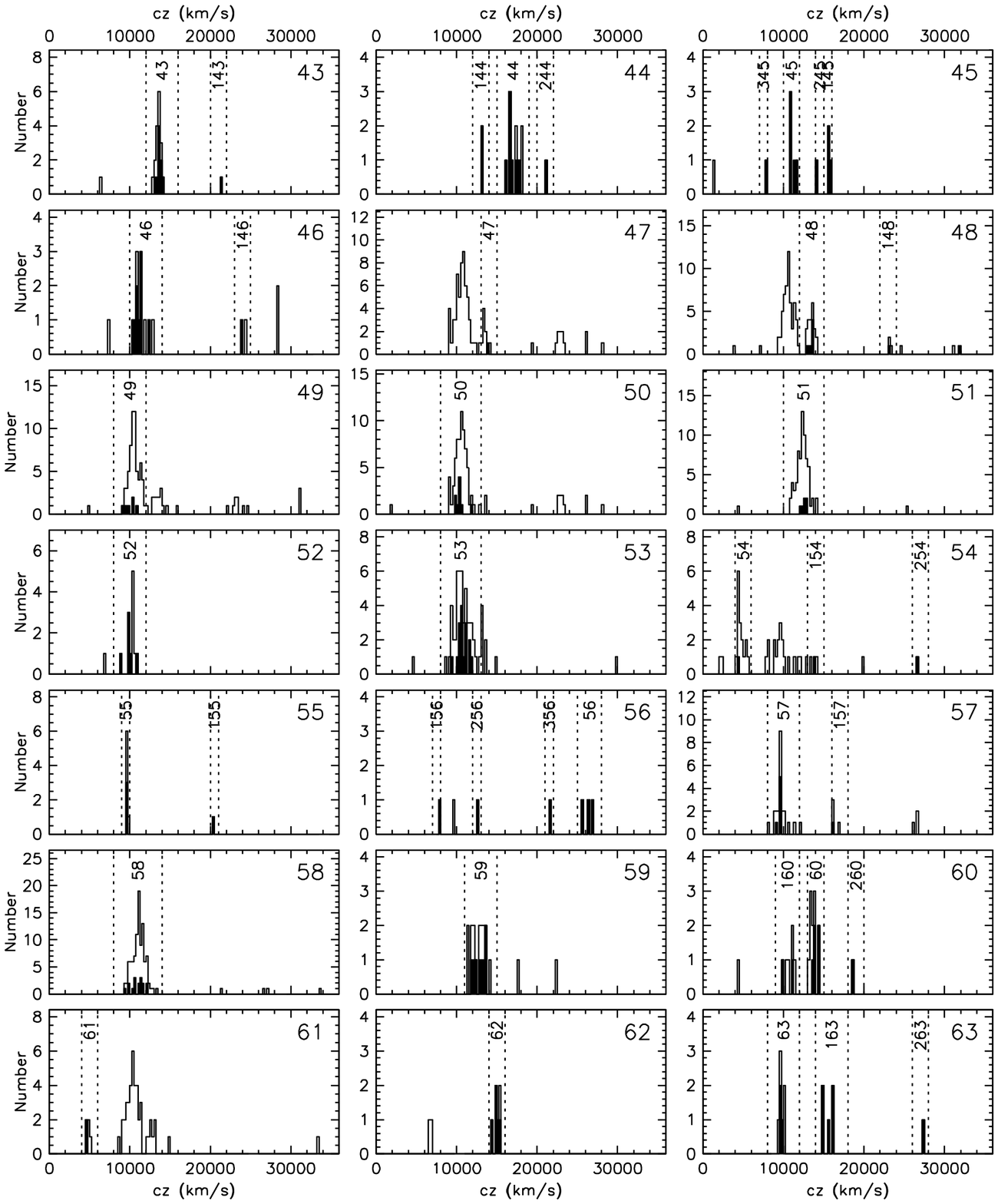}
\caption{{\em (ctd)}}
\vspace*{1cm}
\end{figure*}
\addtocounter{figure}{-1}
\begin{figure*}
\plotfull{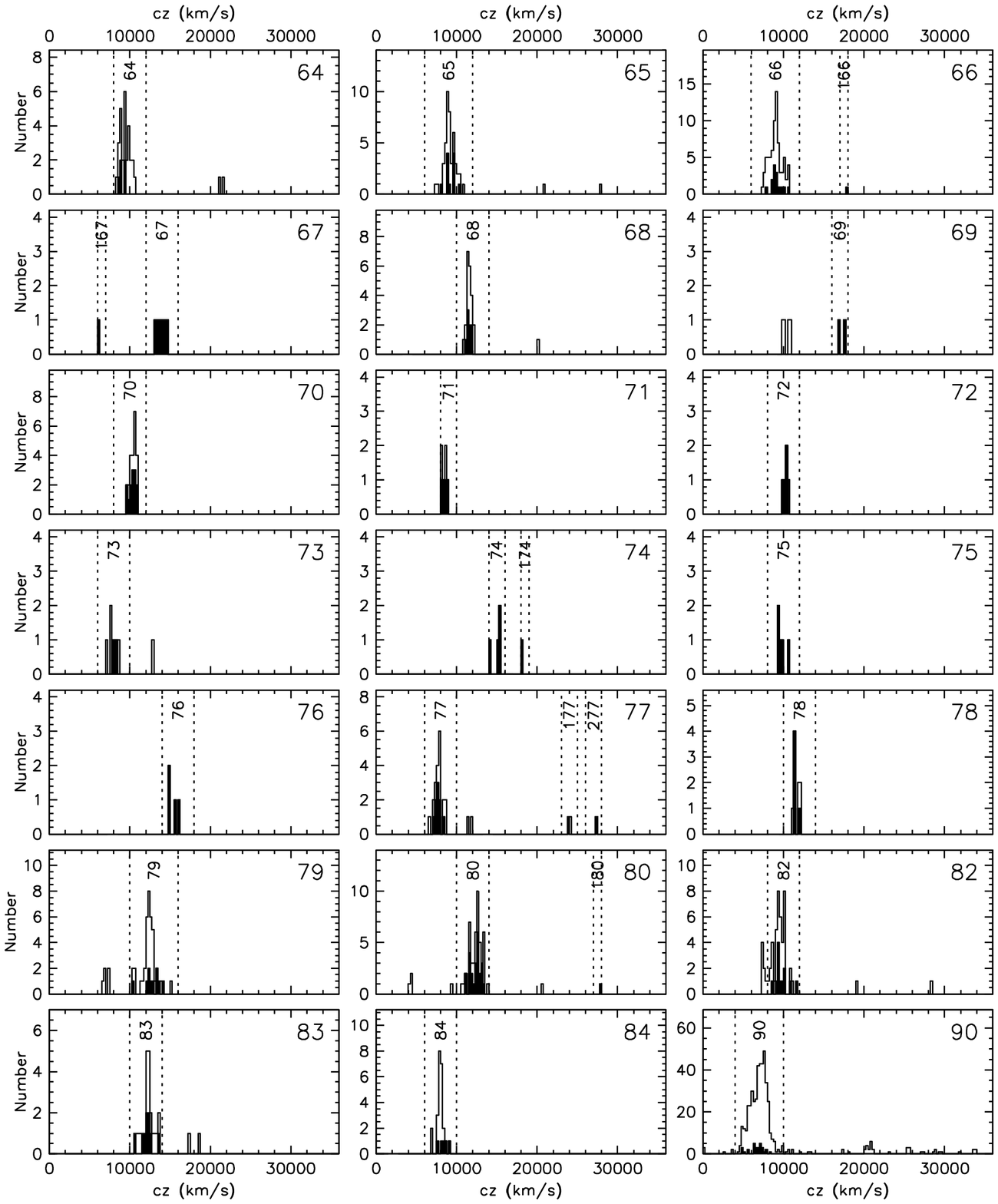}
\caption{{\em (ctd)}}
\vspace*{1cm}
\end{figure*}

\begin{table*}
\centering
\caption{Cluster mean redshifts and velocity dispersions}
\label{tab:cluz}
\renewcommand{\arraystretch}{0.85}
\begin{tabular}{rrrrrrrrrrrrrrrr}
 & \multicolumn{3}{c}{\dotfill EFAR\dotfill}      &
   \multicolumn{3}{c}{\dotfill EFAR+ZCAT\dotfill} & & &
 & \multicolumn{3}{c}{\dotfill EFAR\dotfill}      &
   \multicolumn{3}{c}{\dotfill EFAR+ZCAT\dotfill} \\
~CAN & 
$N$ & \multicolumn{1}{c}{$\langle cz \rangle$} & \multicolumn{1}{c}{$\sigma$} &
$N$ & \multicolumn{1}{c}{$\langle cz \rangle$} & \multicolumn{1}{c}{$\sigma$} & & &
~CAN & 
$N$ & \multicolumn{1}{c}{$\langle cz \rangle$} & \multicolumn{1}{c}{$\sigma$} &
$N$ & \multicolumn{1}{c}{$\langle cz \rangle$} & \multicolumn{1}{c}{$\sigma$}\\
\multicolumn{16}{c}{~} \\
  1 &   6 & 11888 $\pm$   191 &  468 &   6 & 11888 $\pm$   191 &  468 &&& 241 &   2 & 20440 $\pm$  \n94 &  134 &   2 & 20440 $\pm$  \n94 &  134 \\
  2 &   4 & 16332 $\pm$   466 &  931 &  19 & 16454 $\pm$   227 &  991 &&&  42 &   6 &  8783 $\pm$   114 &  280 &  11 &  8856 $\pm$  \n93 &  309 \\
102 &   1 & 22837 $\pm$  \n71 &  --- &  11 & 23340 $\pm$   134 &  444 &&& 142 &   4 & 15863 $\pm$   146 &  292 &   5 & 15917 $\pm$   125 &  280 \\
  3 &  10 & 13168 $\pm$   287 &  907 &  45 & 13280 $\pm$   158 & 1060 &&&  43 &   9 & 13667 $\pm$  \n82 &  246 &  17 & 13536 $\pm$  \n81 &  332 \\
103 &   1 &  4127 $\pm$  \n18 &  --- &   1 &  4127 $\pm$  \n18 &  --- &&& 143 &   1 & 21438 $\pm$  \n20 &  --- &   1 & 21438 $\pm$  \n20 &  --- \\
  4 &   4 & 14090 $\pm$   263 &  527 &   4 & 14090 $\pm$   263 &  527 &&&  44 &   8 & 16950 $\pm$   191 &  539 &  11 & 17184 $\pm$   191 &  633 \\
  5 &   2 & 12082 $\pm$ \n\n8 &   11 &   3 & 11994 $\pm$  \n88 &  153 &&& 144 &   2 & 13190 $\pm$  \n23 &   33 &   2 & 13190 $\pm$  \n23 &   33 \\
105 &   1 & 17154 $\pm$  \n30 &  --- &   1 & 17154 $\pm$  \n30 &  --- &&& 244 &   1 & 21199 $\pm$  \n64 &  --- &   1 & 21199 $\pm$  \n64 &  --- \\
  6 &   7 & 12899 $\pm$   183 &  484 &  11 & 13193 $\pm$   177 &  588 &&&  45 &   5 & 11123 $\pm$   143 &  320 &   5 & 11123 $\pm$   143 &  320 \\
  7 &   7 & 12782 $\pm$   425 & 1123 &   8 & 12625 $\pm$   400 & 1131 &&& 145 &   3 & 15757 $\pm$   109 &  188 &   3 & 15757 $\pm$   109 &  188 \\
107 &   5 & 18296 $\pm$  \n91 &  204 &   5 & 18296 $\pm$  \n91 &  204 &&& 245 &   1 & 14024 $\pm$  \n39 &  --- &   1 & 14024 $\pm$  \n39 &  --- \\
  8 &   5 & 13466 $\pm$  \n49 &  109 &  12 & 13414 $\pm$  \n33 &  116 &&& 345 &   1 &  7789 $\pm$  \n36 &  --- &   1 &  7789 $\pm$  \n36 &  --- \\
108 &   2 &  5276 $\pm$  \n23 &   33 &   5 &  5000 $\pm$   136 &  303 &&&  46 &   9 & 11136 $\pm$   197 &  591 &  12 & 11321 $\pm$   208 &  719 \\
  9 &   2 &  5212 $\pm$   376 &  531 &  19 &  5482 $\pm$   103 &  447 &&& 146 &   1 & 23945 $\pm$  \n21 &  --- &   2 & 24166 $\pm$   221 &  313 \\
109 &   1 &  9599 $\pm$  \n22 &  --- &  15 &  9473 $\pm$   126 &  486 &&&  47 &   1 & 13876 $\pm$  \n26 &  --- &   9 & 13576 $\pm$   113 &  338 \\
 10 &   7 & 15546 $\pm$   270 &  715 &   7 & 15546 $\pm$   270 &  715 &&&  48 &   7 & 13466 $\pm$   114 &  302 &  22 & 13455 $\pm$  \n78 &  364 \\
110 &   1 & 12206 $\pm$  \n50 &  --- &   1 & 12206 $\pm$  \n50 &  --- &&& 148 &   1 & 23099 $\pm$  \n71 &  --- &   3 & 23233 $\pm$   134 &  231 \\
210 &   1 & 20957 $\pm$  \n12 &  --- &   1 & 20957 $\pm$  \n12 &  --- &&&  49 &   7 &  9944 $\pm$   247 &  653 &  64 & 10528 $\pm$  \n80 &  640 \\
 11 &   4 & 14747 $\pm$   163 &  325 &  19 & 14499 $\pm$   179 &  782 &&&  50 &   9 & 10427 $\pm$   193 &  580 &  67 & 10548 $\pm$  \n89 &  730 \\
 12 &  10 & 12342 $\pm$   142 &  448 &  11 & 12315 $\pm$   131 &  435 &&&  51 &   6 & 12593 $\pm$   127 &  312 &  69 & 12353 $\pm$  \n84 &  699 \\
112 &   1 & 20993 $\pm$  \n41 &  --- &   1 & 20993 $\pm$  \n41 &  --- &&&  52 &   6 &  9982 $\pm$   262 &  642 &  12 & 10215 $\pm$   146 &  504 \\
 13 &   8 & 11074 $\pm$   213 &  603 &  10 & 10944 $\pm$   191 &  605 &&&  53 &  16 & 10786 $\pm$   149 &  598 &  49 & 10675 $\pm$   131 &  917 \\
 14 &  10 &  5145 $\pm$   100 &  317 &  59 &  4935 $\pm$  \n61 &  471 &&&  54 &   1 &  4410 $\pm$  \n18 &  --- &  16 &  4699 $\pm$   113 &  452 \\
 15 &   6 & 10725 $\pm$   204 &  499 &   8 & 10655 $\pm$   214 &  606 &&& 154 &   1 & 13830 $\pm$  \n20 &  --- &   3 & 13818 $\pm$   212 &  367 \\
 16 &   9 &  9376 $\pm$   161 &  484 &   9 &  9376 $\pm$   161 &  484 &&& 254 &   1 & 26612 $\pm$  \n73 &  --- &   1 & 26612 $\pm$  \n73 &  --- \\
 17 &   7 & 14417 $\pm$   196 &  519 &   9 & 14355 $\pm$   163 &  490 &&&  55 &   3 &  9635 $\pm$  \n39 &   68 &   7 &  9636 $\pm$  \n29 &   77 \\
 18 &   7 &  8353 $\pm$   169 &  448 &  11 &  8206 $\pm$   232 &  770 &&& 155 &   1 & 20285 $\pm$  \n29 &  --- &   1 & 20285 $\pm$  \n29 &  --- \\
 19 &   2 &  9256 $\pm$   511 &  723 &   2 &  9256 $\pm$   511 &  723 &&&  56 &   3 & 26280 $\pm$   339 &  587 &   3 & 26280 $\pm$   339 &  587 \\
 20 &   8 &  9676 $\pm$   172 &  486 &  14 &  9663 $\pm$   126 &  472 &&& 156 &   1 &  7955 $\pm$  \n15 &  --- &   1 &  7955 $\pm$  \n15 &  --- \\
120 &   1 & 16841 $\pm$  \n15 &  --- &   1 & 16841 $\pm$  \n15 &  --- &&& 256 &   1 & 12730 $\pm$  \n18 &  --- &   1 & 12730 $\pm$  \n18 &  --- \\
 21 &  13 &  7241 $\pm$   210 &  757 &  86 &  7253 $\pm$  \n72 &  663 &&& 356 &   1 & 21558 $\pm$  \n71 &  --- &   1 & 21558 $\pm$  \n71 &  --- \\
 22 &   3 &  8666 $\pm$  \n28 &   48 &   4 &  8649 $\pm$  \n26 &   53 &&&  57 &   6 &  9520 $\pm$   107 &  263 &  22 &  9582 $\pm$   132 &  617 \\
 23 &   9 & 20400 $\pm$   188 &  565 &   9 & 20400 $\pm$   188 &  565 &&& 157 &   1 & 16075 $\pm$  \n45 &  --- &   4 & 16237 $\pm$   192 &  384 \\
123 &   2 & 11852 $\pm$  \n34 &   49 &   2 & 11852 $\pm$  \n34 &   49 &&&  58 &  16 & 10943 $\pm$   197 &  789 &  99 & 11106 $\pm$  \n79 &  781 \\
 24 &   6 &  9651 $\pm$   204 &  499 & 137 &  9854 $\pm$  \n66 &  776 &&&  59 &   8 & 12864 $\pm$   242 &  683 &  16 & 12693 $\pm$   215 &  862 \\
 25 &  10 & 10999 $\pm$   204 &  646 &  12 & 11021 $\pm$   172 &  596 &&&  60 &   5 & 13993 $\pm$   142 &  318 &  11 & 13707 $\pm$   119 &  393 \\
 26 &   3 &  8677 $\pm$   297 &  514 &  12 &  8937 $\pm$   189 &  653 &&& 160 &   2 & 10474 $\pm$   651 &  921 &   6 & 10730 $\pm$   223 &  547 \\
 27 &   5 &  4540 $\pm$   188 &  420 &  30 &  4436 $\pm$  \n65 &  355 &&& 260 &   1 & 18545 $\pm$  \n26 &  --- &   1 & 18545 $\pm$  \n26 &  --- \\
 28 &   5 & 10030 $\pm$   363 &  813 &  17 &  9884 $\pm$   208 &  857 &&&  61 &   2 &  4742 $\pm$  \n16 &   11 &   5 &  4878 $\pm$  \n90 &  202 \\
128 &   1 & 16920 $\pm$  \n27 &  --- &   1 & 16920 $\pm$  \n27 &  --- &&&  62 &   5 & 14918 $\pm$   167 &  373 &   6 & 14991 $\pm$   154 &  378 \\
 29 &   5 &  9492 $\pm$   339 &  759 &  18 &  9528 $\pm$   181 &  769 &&&  63 &   3 &  9699 $\pm$  \n92 &  159 &   7 &  9777 $\pm$   116 &  306 \\
129 &   1 & 16483 $\pm$  \n18 &  --- &   1 & 16483 $\pm$  \n18 &  --- &&& 163 &   5 & 15563 $\pm$   273 &  610 &   5 & 15563 $\pm$   273 &  610 \\
 30 &   1 & 12711 $\pm$  \n19 &  --- &   2 & 12576 $\pm$   135 &  191 &&& 263 &   1 & 27486 $\pm$  \n71 &  --- &   1 & 27486 $\pm$  \n71 &  --- \\
130 &   1 & 16760 $\pm$  \n14 &  --- &   1 & 16760 $\pm$  \n14 &  --- &&&  64 &   5 &  9062 $\pm$   154 &  345 &  28 &  9423 $\pm$   113 &  598 \\
 31 &   3 &  7288 $\pm$  \n47 &   81 &   5 &  7289 $\pm$  \n31 &   69 &&&  65 &  11 &  9223 $\pm$   171 &  568 &  46 &  9137 $\pm$  \n97 &  659 \\
131 &   3 &  9384 $\pm$   171 &  296 &   6 &  9299 $\pm$   102 &  250 &&&  66 &  15 &  9156 $\pm$   178 &  689 &  73 &  9014 $\pm$  \n93 &  796 \\
 32 &   5 & 16416 $\pm$  \n94 &  210 &   5 & 16416 $\pm$  \n94 &  210 &&& 166 &   1 & 17795 $\pm$  \n44 &  --- &   1 & 17795 $\pm$  \n44 &  --- \\
132 &   3 & 13499 $\pm$   144 &  250 &   3 & 13499 $\pm$   144 &  250 &&&  67 &   7 & 13876 $\pm$   215 &  570 &   7 & 13876 $\pm$   215 &  570 \\
232 &   1 &  8926 $\pm$ \n\n9 &  --- &   1 &  8926 $\pm$ \n\n9 &  --- &&& 167 &   1 &  6086 $\pm$  \n12 &  --- &   1 &  6086 $\pm$  \n12 &  --- \\
332 &   1 &  4758 $\pm$ \n\n9 &  --- &   1 &  4758 $\pm$ \n\n9 &  --- &&&  68 &   8 & 11514 $\pm$  \n97 &  273 &  22 & 11547 $\pm$  \n67 &  316 \\
 33 &   7 & 11652 $\pm$   340 &  899 &   7 & 11652 $\pm$   340 &  899 &&&  69 &   2 & 17344 $\pm$   403 &  570 &   2 & 17344 $\pm$   403 &  570 \\
 34 &   8 & 14498 $\pm$   265 &  749 &   9 & 14488 $\pm$   234 &  702 &&&  70 &  13 & 10320 $\pm$   109 &  392 &  23 & 10396 $\pm$  \n78 &  376 \\
 35 &  27 & 11834 $\pm$   133 &  694 &  66 & 11866 $\pm$  \n78 &  630 &&&  71 &   4 &  8465 $\pm$   147 &  294 &   6 &  8442 $\pm$   113 &  278 \\
 36 &   6 & 12861 $\pm$   309 &  757 &  69 & 12732 $\pm$   101 &  837 &&&  72 &   5 & 10311 $\pm$   136 &  305 &   5 & 10311 $\pm$   136 &  305 \\
136 &   2 &  8764 $\pm$   234 &  331 &   2 &  8764 $\pm$   234 &  331 &&&  73 &   3 &  8121 $\pm$   176 &  305 &   7 &  7935 $\pm$   194 &  514 \\
 37 &   6 &  8888 $\pm$  \n81 &  199 &   8 &  8866 $\pm$  \n82 &  231 &&&  74 &   4 & 14965 $\pm$   267 &  534 &   4 & 14965 $\pm$   267 &  534 \\
137 &   1 & 11399 $\pm$  \n18 &  --- &   1 & 11399 $\pm$  \n18 &  --- &&& 174 &   1 & 18187 $\pm$  \n36 &  --- &   1 & 18187 $\pm$  \n36 &  --- \\
 38 &   4 & 12788 $\pm$   280 &  559 &   6 & 12695 $\pm$   252 &  616 &&&  75 &   5 &  9784 $\pm$   233 &  521 &   5 &  9784 $\pm$   233 &  521 \\
138 &   1 & 15405 $\pm$  \n16 &  --- &   1 & 15405 $\pm$  \n16 &  --- &&&  76 &   4 & 15365 $\pm$   320 &  640 &   4 & 15365 $\pm$   320 &  640 \\
238 &   1 & 17845 $\pm$  \n27 &  --- &   1 & 17845 $\pm$  \n27 &  --- &&&  77 &   9 &  7639 $\pm$   124 &  373 &  20 &  7728 $\pm$   112 &  502 \\
338 &   1 & 10873 $\pm$  \n35 &  --- &   5 & 10438 $\pm$   236 &  528 &&& 177 &   1 & 23926 $\pm$  \n12 &  --- &   2 & 23992 $\pm$  \n66 &   93 \\
 39 &  12 &  8882 $\pm$  \n69 &  238 &  18 &  8832 $\pm$  \n61 &  257 &&& 277 &   1 & 27282 $\pm$  \n13 &  --- &   1 & 27282 $\pm$  \n13 &  --- \\
139 &   3 & 19044 $\pm$   183 &  317 &   4 & 19076 $\pm$   133 &  266 &&&  78 &   6 & 11504 $\pm$   133 &  327 &   9 & 11571 $\pm$   136 &  408 \\
239 &   1 & 17438 $\pm$  \n55 &  --- &   2 & 17443 $\pm$ \n\n5 &    7 &&&  79 &  10 & 12692 $\pm$   328 & 1036 &  40 & 12441 $\pm$   149 &  944 \\
339 &   1 & 15651 $\pm$  \n30 &  --- &   1 & 15651 $\pm$  \n30 &  --- &&&  80 &  24 & 12331 $\pm$   133 &  649 &  48 & 12399 $\pm$   107 &  740 \\
 40 &   3 & 11612 $\pm$   217 &  376 &   4 & 11846 $\pm$   280 &  560 &&& 180 &   1 & 27823 $\pm$ \n\n8 &  --- &   1 & 27823 $\pm$ \n\n8 &  --- \\
140 &   2 &  9806 $\pm$   160 &  226 &   2 &  9806 $\pm$   160 &  226 &&&  82 &  12 &  9771 $\pm$   238 &  824 &  43 &  9573 $\pm$   113 &  741 \\
240 &   1 & 17291 $\pm$  \n26 &  --- &   1 & 17291 $\pm$  \n26 &  --- &&&  83 &   9 & 12157 $\pm$   253 &  759 &  23 & 12252 $\pm$   156 &  748 \\
340 &   1 & 28973 $\pm$  \n28 &  --- &   1 & 28973 $\pm$  \n28 &  --- &&&  84 &   5 &  8345 $\pm$   224 &  500 &  24 &  7962 $\pm$  \n90 &  442 \\
 41 &   5 &  8973 $\pm$   135 &  303 &   7 &  8848 $\pm$   127 &  335 &&&  90 &  29 &  6663 $\pm$   172 &  924 & 435 &  6942 $\pm$  \n50 & 1034 \\
141 &   2 & 13685 $\pm$   405 &  573 &   3 & 13794 $\pm$   258 &  447 \\
\end{tabular}
\end{table*}

The results of this process are shown in Figure~\ref{fig:cluz}, which
shows the redshift distributions of galaxies within 3\Mpc\ around each
of the nominal EFAR clusters (labelled by their cluster ID number, CID;
see Paper~I) and the adopted groupings in redshift space. Note that
CID=81 (A2593-S) does not appear since it was merged with CID=80
(A2593-N)---see below. Each EFAR galaxy was assigned to one of these
groupings and given a cluster assignment number (CAN), listed in
Table~\ref{tab:galtab}. The main grouping along the line of sight has a
CAN which is simply the original two-digit CID; other groupings have
CANs with a distinguishing third leading digit. The groupings (which we
will hereafter call clusters regardless of their size) are labelled by
their CANs in Figure~\ref{fig:cluz}, which also shows the boundaries of
each cluster in redshift space. The last two digits of each galaxy's CAN
is its CID, apart from 41 galaxies which were reassigned to other
neighbouring clusters: two galaxies in CID=33 were reassigned to CAN=34
(GINs 254, 255); two galaxies in CID=34 were reassigned to CAN=33 (GINs
263, 264); five galaxies in CID=35 were reassigned to CAN=36 (GINs 270,
274, 275, 281, 282); fourteen galaxies in CID=36 were reassigned to
CAN=35 (GINS 285--292, 295--297, 299--301), one galaxy in CID=47 was
reassigned to CAN=50 (GIN 406); three galaxies in CID=59 and two in
CID=61 were reassigned to CAN=53 (GINs 514, 517, 527, 536, 537); five
galaxies with CID=69 were reassigned to CAN=70 (GINs 617, 618, 619, 622,
623); and all seven galaxies with CID=81 were reassigned to CAN=80 (GINs
709--715).

Table~\ref{tab:cluz} lists, for each CAN, the number of EFAR galaxies,
the number of EFAR+ZCAT galaxies, and the mean redshift, its standard
error (taken to the error in the redshift for clusters with only one
member) and the velocity dispersion. These quantities are computed both
from the EFAR sample and from the EFAR+ZCAT sample. In many of the
clusters the EFAR sample is greatly supplemented by the ZCAT galaxies,
leading to much-improved estimates of the mean cluster redshift: using
EFAR galaxies only the median uncertainty in the mean cluster redshift
(for clusters with more than one member) is 177\kms; with EFAR+ZCAT
galaxies the median uncertainty is reduced to 133\kms.

\section{CONCLUSIONS}
\label{sec:conclude}

We have described the observations, reductions, and analysis of 1319
spectra of 714 early-type galaxies studied as part of the EFAR project.
We have obtained redshifts for 706 galaxies, velocity dispersions and
\mgb\ linestrengths for 676 galaxies, and \mgtwo\ linestrengths for 582
galaxies. Although obtained in 33 observing runs spanning seven years
and 10 different telescopes, we have applied uniform procedures to
derive the spectroscopic parameters and brought all the measurements of
each parameter onto a standard system which we ensure is internally
consistent through comparisons of the large numbers of repeat
measurements, and externally consistent through comparisons with
published data. We have performed detailed simulations to estimate
measurement errors and calibrated these error estimates using the repeat
observations.

The fully-corrected measurements of each parameter from the individual
spectra are given in Table~\ref{tab:spectab}; the final parameters for
706 galaxies, computed as the appropriately-weighted means of the
individual measurements, are listed in Table~\ref{tab:galtab}. The
median estimated errors in the combined measurements (including
measurement errors and run correction uncertainties) are $\Delta
cz$=20\kms, $\Delta\sigma/\sigma$=9.1\% (\ie\
$\Delta\log\sigma$=0.040~dex), $\Delta\mgb/\mgb$=7.2\% (\ie\
$\Delta\mgbp$=0.013~mag) and $\Delta\mgtwo$=0.015~mag. Comparisons with
redshifts and dispersions from the literature show no systematic errors.
The linestrengths required only small zeropoint corrections to bring
them onto the Lick system.

We have assigned galaxies to physical clusters (as opposed to apparent
projected clusters) by examining the line-of-sight velocity
distributions based on EFAR and ZCAT redshifts, together with the
projected distributions on the sky. We derive mean redshifts for these
physical clusters, which will be used in estimating distances and
peculiar velocities, and also velocity dispersions, which will be used
to test for trends in the galaxy population with cluster mass or local
environment.

The results presented here comprise the largest single set of velocity
dispersions and linestrengths for early-type galaxies published to
date. These data will be used in combination with the sample selection
criteria of Wegner \etal\ (1996, Paper~I) and the photometric data of
Saglia \etal\ (1997, Paper~III) to analyse the properties and peculiar
motions of early-type galaxies in the two distant regions studied by the
EFAR project.

\section*{Acknowledgements}

We gratefully acknowledge all the observatories which supported this
project: MMC, RKM, RLD and DB were Visiting Astronomers at Kitt Peak
National Observatory, while GB, RLD and RKM were Visiting Astronomers at
Cerro Tololo Inter-American Observatory---both observatories are
operated by AURA, Inc.\ for the National Science Foundation; GW and DB
used MDM Observatory, operated by the University of Michigan, Dartmouth
College and the Massachusetts Institute of Technology; DB and RKM used
the Multiple Mirror Telescope, jointly operated by the Smithsonian
Astrophysical Observatory and Steward Observatory; RPS used facilities
at Calar Alto (Centro Astrofisico Hispano Alemano) and La Silla (ESO);
MMC observed at Siding Spring (MSSSO); MMC, RLD, RPS and GB used the
telescopes of La Palma Observatory. We thank the many support staff at
these observatories who assisted us with our observations. We thank
S.Sakai for doing one observing run. We also thank the SMAC team for
providing comparison data prior to publication, and Mike Hudson for
helpful discussions.

We also gratefully acknowledge the financial support provided by various
funding agencies: GW was supported by the SERC and Wadham College during
a year's stay in Oxford, and by the Alexander von Humboldt-Stiftung
during a visit to the Ruhr-Universit\"{a}t in Bochum; MMC acknowledges
the support of a Lindemann Fellowship, DIST Collaborative Research
Grants and an Australian Academy of Science/Royal Society Exchange
Program Fellowship; RPS was supported by DFG grants SFB 318 and 375.
This work was partially supported by NSF Grant AST90-16930 to DB,
AST90-17048 and AST93-47714 to GW, and AST90-20864 to RKM. The entire
collaboration benefitted from NATO Collaborative Research Grant 900159
and from the hospitality and financial support of Dartmouth College,
Oxford University, the University of Durham and Arizona State
University. Support was also received from PPARC visitors grants to
Oxford and Durham Universities and a PPARC rolling grant `Extragalactic
Astronomy and Cosmology in Durham 1994-98'.

\appendix

\section*{APPENDIX A: OBSERVING DETAILS}

This appendix gives further details of the instrumental configurations
used on different telescopes.

MDM 2.4m: The Mark IIIa spectrograph, with a 1.87~arcsec wide slit,
was used for all runs up to the end of 1988; from 1989 this was
replaced by the Mark IIIb spectrograph, which is identical except that
a 1.68~arcsec slit was used (except for run 113, when the slit width
was 2.36~arcsec). For runs 101--103 a 600~lines/mm grism blazed at
4600\AA\ was used; for all subsequent runs, a 600~lines/mm grism
blazed at 5700\AA\ was employed. The slit was usually oriented
N--S. Two-pixel binning perpendicular to the dispersion direction was
employed to lower the readout noise.
 
KPNO 4m: The RC spectrograph and grating KPC-17B (527~lines/mm) were
used with the UV Fast Camera and the TI2 CCD.

KPNO 2.1m: The Gold spectrograph/camera and  grating \#240
(500~lines/mm) were  used with the TI5 CCD.
 
WHT 4.2m: The blue arm of the ISIS spectrograph was used with the
CCD-IPCS imaging photon counting system. Most objects were observed
using the R600B grating (600~lines/mm), but one object (J26~A,
GIN=648) was observed with the R300B grating (300~lines/mm).
 
INT 2.5m: The Intermediate Dispersion Spectrograph (IDS) and R632V
grating (632~lines/mm) were used with the 235mm camera for all runs.
                                   
SSO 2.3m: Both runs used the blue arm of the Double Beam Spectrograph
(DBS) with a 600~lines/mm grating. Run 130 used the Photon Counting
Array (PCA), while run 132 used a Loral CCD.
 
MMT Blue: The `Big Blue' spectrograph was employed with a 300~lines/mm
grating (blazed at 4800\AA in first order) and the Reticon
detector. The MMT image stacker gave two 2.5~arcsec circular apertures
separated by 36~arcsec.

MMT Red: The MMT Red Channel was used with a 600~lines/mm grating
(blazed at 4800\AA) and the $800\times800$ TI CCD binned by two pixels
perpendicular to the dispersion. The slit was $1.5\times180$~arcsec,
but heavily vignetted in the outer 30~arcsec in one direction.

Calar Alto 2.2m: The Cassegrain Boller \& Chivens slit spectrograph
with grating \#7 (60~\AA/mm) was used in combination with the TEK\#6
CCD.
 
ESO 3.6m: The MEFOS fibre feed and the Boller \& Chivens spectrograph
were used. MEFOS has 58 2.6~arcsec diameter fibres (29 for targets and
29 for sky) positioned within a 1~degree diameter field at prime
focus. The detector was a Tektronix TK512CB CCD (ESO\#32).

CTIO 4m: The ARGUS 24-object fibre spectrograph was used. ARGUS has a
50~arcmin field at the f/2.8 prime focus. Each of the 24 arms holds
two 1.9~arcsec diameter fibres which lie 36~arcsec apart on the sky;
one arm is positioned on the target and the other on sky. The fibres
feed a thermally and mechanically isolated bench spectrograph with a
510mm focal length Schmidt blue collimator and a 229mm focal length
Schmidt camera. A Reticon~II $1200\times400$ CCD detector was used
with grating KPGL~\#3.

\end{document}